%% file: paper.tex
\documentclass[10pt,journal,compsoc]{IEEEtran}
\ifCLASSOPTIONcompsoc
  \usepackage[nocompress]{cite}
\else
  \usepackage{cite}
\fi
\ifCLASSINFOpdf
\else
\fi

\usepackage{amssymb}
\usepackage{hyperref}
\usepackage[dvipdfmx]{graphicx}
\usepackage{bmpsize}
\usepackage{mathtools}
\usepackage[table]{xcolor}
\usepackage{colortbl}
\usepackage{algorithmic}
\usepackage{algorithm}

\usepackage{booktabs}
\usepackage{multirow}
 \usepackage{enumitem}
\usepackage{listings}
\usepackage{xcolor}
\usepackage{lstautogobble}
\usepackage{amsfonts}
\usepackage{amssymb}
\definecolor{codegreen}{rgb}{0,0.6,0}
\definecolor{codegray}{rgb}{0.5,0.5,0.5}
\definecolor{codepurple}{rgb}{0.58,0,0.82}
\definecolor{backcolour}{rgb}{0.95,0.95,0.92}

\usepackage{pgfplotstable}
\usepackage{pgfplots}
\usepackage{subcaption}
\RequirePackage{subcaption}
\usepackage{makecell}
\lstdefinestyle{mystyle}{
    backgroundcolor=\color{backcolour},
    commentstyle=\color{codegreen},
    keywordstyle=\color{magenta},
    numberstyle=\tiny\color{codegray},
    stringstyle=\color{codepurple},
    basicstyle=\ttfamily\footnotesize,
    breakatwhitespace=false,
    breaklines=true,
    captionpos=b,
    keepspaces=true,
    numbers=left,
    numbersep=5pt,
    showspaces=false,
    showstringspaces=false,
    showtabs=false,
    tabsize=2
}

\definecolor{darkblue}{RGB}{18,22,46}
\definecolor{mediumblue}{RGB}{51,70,99}
\definecolor{lightblue}{RGB}{94,125,141}
\definecolor{verylightblue}{RGB}{185,196,189}

\definecolor{darkred}{RGB}{98,24,13}
\definecolor{orange}{RGB}{184,101,26}
\definecolor{lightyellow}{RGB}{194,171,85}
\definecolor{verylightyellow}{RGB}{200,197,132}

\pgfplotscreateplotcyclelist{mycolorlist}{%
lightblue,mark=*\\
mediumblue,mark=*\\
darkred,mark=*\\
orange,mark=*\\
lightyellow,mark=*\\
verylightyellow,mark=*\\
}

\usepackage{tikz}
\usepackage{pgfplots}
\pgfplotsset{compat=1.16}
\usepgfplotslibrary{groupplots}
\pgfplotsset{
  every axis/.append style={
    no markers,
    grid=major,
    grid style={dashed},
    legend style={font=\tiny},
    ylabel style={font=\scriptsize},
    xlabel style={font=\scriptsize},
  },
  every axis plot/.append style={line width=1.2pt, line join=round},
  every axis legend/.append style={legend columns=1},
  group/group size=3 by 1,
  every x tick label/.append style={alias=XTick,inner xsep=0pt},
  every x tick scale label/.style={at=(XTick.base east),anchor=base west}
}

\definecolor{col1}{RGB}{53, 110, 175}
\definecolor{col2}{RGB}{204, 42, 42}
\definecolor{col3}{RGB}{255, 175, 35}
\definecolor{col4}{RGB}{79, 162, 46}
\definecolor{col5}{RGB}{97, 97, 97}
\definecolor{col6}{RGB}{103, 63, 153}
\definecolor{col7}{RGB}{0, 0, 0}
\definecolor{col8}{RGB}{123, 63, 0}

\tikzset{
  curve1/.style={col1},
  curve2/.style={col2},
  curve3/.style={col4},
  curve4/.style={col3},
  curve5/.style={col6},
  curve6/.style={cyan},
  curve7/.style={col5, dashdotted},
  curve8/.style={col7, dashed},
  curve9/.style={col8, densely dotted},
  curve10/.style={teal, densely dotted},
  curve11/.style={lime},
  curve12/.style={orange},
}

\input{notations.tex}

\graphicspath{{figures/}{img/}}

\begin{document}
%
\title{TTK is Getting MPI-Ready}
\title{\julien{Distributing Topological Analysis Pipelines}}
\title{\major{\ourNewTitle}}
%
%
%
%

\author{
E. Le Guillou,
M. Will,
P. Guillou,
J. Lukasczyk,
P. Fortin,
C. Garth,
J. Tierny
\IEEEcompsocitemizethanks{
\IEEEcompsocthanksitem E. Le Guillou is with the CNRS, Sorbonne Université and
University of Lille.
E-mail:
\href{mailto:eve.le_guillou@sorbonne-universite.fr}{eve.le\_guillou@sorbonne-
universite.fr}
\IEEEcompsocthanksitem M. Will, J. Lukasczyk and C. Garth are with RPTU
Kaiserslautern-Landau.
E-mails:
\href{mailto:mswill@rptu.de,
lukasczyk@rptu.de, garth@rptu.de}{\{mswill,lukasczyk,garth\}@rptu.de}
\IEEEcompsocthanksitem P. Guillou and J. Tierny are with the CNRS and Sorbonne
Université.
E-mails:
\href{mailto:pierre.guillou@sorbonne-universite.fr, julien.tierny@sorbonne-
universite.fr}{\{firstname.lastname\}@sorbonne-universite.fr}
\IEEEcompsocthanksitem P. Fortin is with \pierre{Univ. Lille, CNRS, Centrale
Lille, UMR 9189 CRIStAL, F-59000 Lille, France.
E-mail: \href{mailto:pierre.fortin@univ-lille.fr}{pierre.fortin@univ-lille.fr}}
}

\thanks{Manuscript received May 21, 2021; revised May 11, 2021.}}

%
%

\markboth{Journal of \LaTeX\ Class Files,~Vol.~14, No.~8, May~2021}%
{Shell \MakeLowercase{\textit{et al.}}: Bare Demo of IEEEtran.cls for Computer Society Journals}
%



\IEEEtitleabstractindextext{%

\begin{abstract}
\major{This system paper documents the technical foundations for the extension of the \emph{Topology ToolKit} (TTK) to distributed-memory parallelism with the \emph{Message Passing Interface} (MPI).}
While several recent papers introduced topology-based approaches for
\pierre{distributed-memory}
environments, these were reporting experiments obtained with tailored,
mono-algorithm implementations.
In contrast, we
\julien{describe in this paper}
a \major{versatile approach (supporting both triangulated domains and regular
grids) for the support of}
\julien{topological analysis \emph{pipelines}, i.e.\major{,} a sequence of
topological algorithms interacting together, possibly on distinct numbers of
processes.}
%
\julien{While developing this \major{extension},}
we faced several algorithmic and software engineering
challenges, which we
document in this paper.
\major{Specifically, we}
describe
\julien{an MPI extension}
of TTK's
data structure for triangulation representation and traversal, a
central component
to the global performance and
\majorRevision{generality}
of TTK's topological implementations.
We also introduce an intermediate interface between TTK and MPI,
both at the \julien{global} pipeline level, and at the fine-grain algorithmic level.
We provide a taxonomy for the \pierrre{distributed-memory topological algorithms}
supported by TTK, depending on their communication needs and provide examples of
\pierre{hybrid MPI+thread}
\pierrre{parallelizations.}
\julien{Detailed performance analyses show that parallel efficiencies range from $20\%$ to $80\%$ (depending on the algorithms), and that
the MPI-specific preconditioning introduced by our framework induces a negligible computation time overhead.}
We illustrate the new \pierre{distributed-memory} capabilities of TTK with an example of
advanced analysis pipeline, combining multiple algorithms,
run on
\julien{the largest publicly available dataset we have found ($120$ billion vertices)}
on a \julien{standard} \pierre{cluster} with
\eve{$64$}
nodes (for a total of
\majorRevision{$1536$}
cores).
Finally, we provide a roadmap for the completion of TTK's MPI
\julien{extension,}
along with
generic recommendations for each algorithm communication category.
\end{abstract}


\begin{IEEEkeywords}
Topological data analysis, high-performance computing, \pierre{distributed-memory} algorithms.
\end{IEEEkeywords}}

\maketitle

\IEEEdisplaynontitleabstractindextext

%
\IEEEpeerreviewmaketitle

\IEEEraisesectionheading{\section{Introduction}\label{sec:introduction}}

\input{introduction.tex}

\input{background.tex}

\input{setup.tex}

\input{triangulation.tex}
\input{distributing_ttk.tex}
\input{examples.tex}
\input{results.tex}

\input{conclusion.tex}

\ifCLASSOPTIONcaptionsoff
  \newpage
\fi



%
%
%

\section*{Acknowledgments}
\footnotesize{
This work is partially supported by the
European Commission grant
ERC-2019-COG
\emph{``TORI''} (ref. 863464,
\url{https://erc-tori.github.io/}).}


\bibliographystyle{abbrv-doi}
\bibliography{paper}

\begin{IEEEbiography}[{\includegraphics[width=1in,height=1.25in,clip,
keepaspectratio]{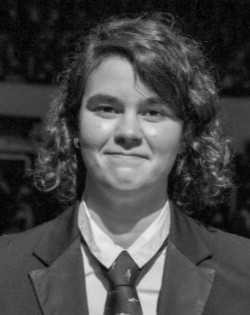}}]{Eve Le Guillou}
is a Ph.D. student at Sorbonne Université. She received a M.S. degree in 2020 in Computer Science
from Cranfield University as well as an engineering degree in 2021 from École Centrale de Lille.
She is an active contributor to the Topology ToolKit (TTK), an open source library for
topological data analysis. Her notable contributions to TTK include the port to MPI
of TTK's data structure and of several algorithms.
\end{IEEEbiography}

\begin{IEEEbiography}[{\includegraphics[width=1in,height=1.25in,clip,
  keepaspectratio]{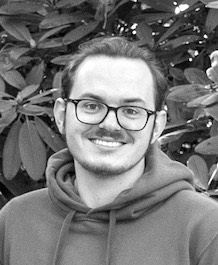}}]{Michael Will}
is a Ph.D. student at RPTU Kaiserslautern-Landau. He received a M.S. degree in Computer Science
from the Technical University of Kaiserslautern in 2021.
He is an active contributor to TTK's MPI section. His research interest include topological data analysis,
especially using high performance computing on large-scale data.
\end{IEEEbiography}

\begin{IEEEbiography}[{\includegraphics[width=1in,height=1.25in,clip,
keepaspectratio]{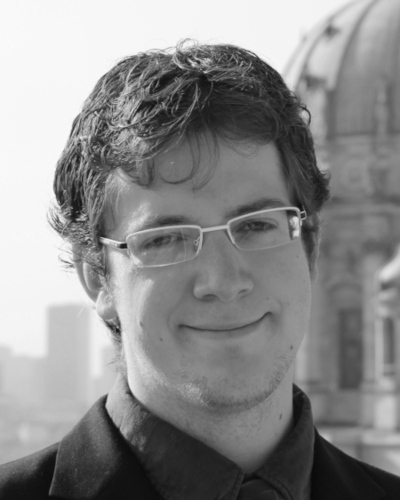}}]{Pierre Guillou}
is a research engineer at Sorbonne Université. After graduating from
MINES ParisTech, a top French engineering school in 2013, he received
his Ph.D., also from MINES ParisTech, in 2016. His Ph.D. work revolved
around parallel image processing algorithms for embedded
accelerators. Since 2019, he has been an active contributor to TTK and
the author of many modules created for the VESTEC and TORI projects.
\end{IEEEbiography}




\begin{IEEEbiography}[{\includegraphics[width=1in,height=1.25in,
        clip,keepaspectratio]{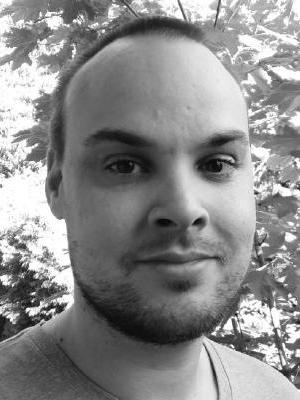}}]{Jonas Lukasczyk}
received his Ph.D. degree from the Visual Information Analysis Group, Technische
Universitat Kaiserslautern, Germany, where he also studied applied computer
science and mathematics. His recent work focuses on topology-based
characterization of features and their evolution in large-scale simulations.
\end{IEEEbiography}

\begin{IEEEbiography}[{\includegraphics[width=1in,height=1.25in,
        clip,keepaspectratio]{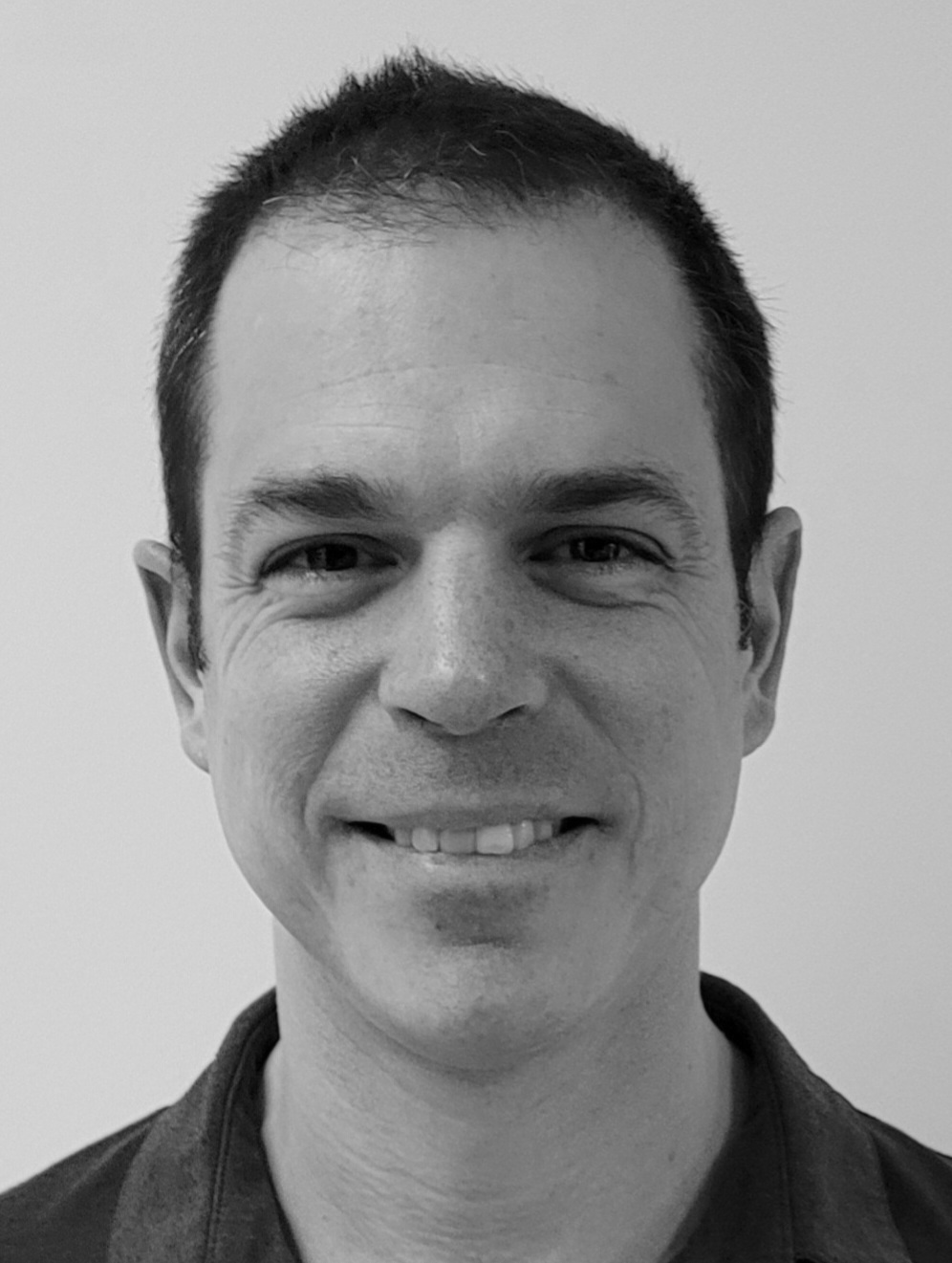}}]{Pierre Fortin}
  received his Ph.D. degree in Computer Science from the University
  of Bordeaux in 2006, and his Habilitation degree (HDR)
  from Sorbonne University in
  2018. He has been Assistant Professor from 2007 to 2020 at Sorbonne
  University, and he is Associate Professor
  at University of Lille since 2020. His research interests include parallel
  algorithmic and high performance scientific computing.
\end{IEEEbiography}

\begin{IEEEbiography}[{\includegraphics[width=1in,height=1.25in,
        clip,keepaspectratio]{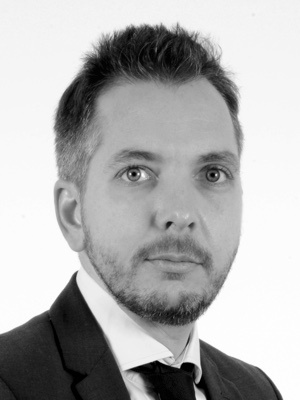}}]{Christoph Garth}
received the PhD degree in
computer science from Technische Universität
(TU) Kaiserslautern in 2007. After four years as
a postdoctoral
researcher with the University of
California, Davis, he rejoined TU Kaiserslautern
where he is currently a full professor of computer
science. His research interests include large scale data analysis and
visualization, in situ visualization, topology-based methods,
and interdisciplinary applications of visualization.
\end{IEEEbiography}

\begin{IEEEbiography}[{\includegraphics[width=1in,height=1.25in,clip,
keepaspectratio]{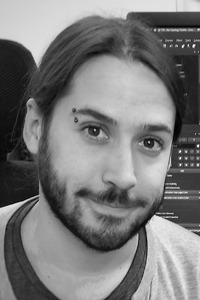}}]{Julien Tierny}
received the Ph.D. degree in Computer Science from the University
of Lille in 2008.
He is
a CNRS research director at
Sorbonne University. Prior to his CNRS tenure, he held a
Fulbright fellowship
and was a post-doctoral
researcher at the
University
of Utah.
His research expertise lies in topological methods for data analysis
and visualization.
He is the founder and lead developer of the Topology ToolKit
(TTK), an open source library for topological data analysis.
\end{IEEEbiography}

%
%
%






\clearpage
\includegraphics{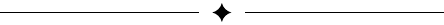}
\input{appendixBody}


\end{document}

%% file: notations.tex
\newcommand{\discreteVectorField}{\mathcal{V}}
\newcommand{\numberProcesses}{n_p}

\newcommand{\domain}{\mathcal{M}}

\newcommand{\sublevelset}[1]{#1^{-1}_{-\infty}}

\newcommand{\Star}{St}
\newcommand{\Link}{Lk}
\newcommand{\simplex}{\sigma}

\newcommand{\Index}{\mathcal{I}}

\newcommand{\forwardIntegralLine}{\mathcal{L}^+}
\newcommand{\backwardIntegralLine}{\mathcal{L}^-}

\newcommand{\boundingBox}{\mathcal{B}}

\newcommand{\julien}[1]{\textcolor{black}{#1}}
\definecolor{orange}{RGB}{184,101,26}
\newcommand{\eve}[1]{\textcolor{black}{#1}}
\newcommand{\pierre}[1]{\textcolor{black}{#1}}
\newcommand{\pierrre}[1]{\textcolor{black}{#1}}
\newcommand{\toDelete}[1]{\textcolor{black}{#1}}

\newcommand{\majorRevision}[1]{\textcolor{black}{#1}}
\newcommand{\major}[1]{\majorRevision{#1}}
\newcommand{\discuss}[1]{\textcolor{black}{#1}}

\newcommand{\ourNewTitle}{A Generic Software Framework for\\Distributed 
Topological Analysis Pipelines}
\renewcommand{\ourNewTitle}{TTK is Getting MPI-Ready}

%% file: introduction.tex

\major{Modern datasets} are constantly growing in size,
\major{due}
to
the continuous improvements of acquisition technologies \major{and}
computational
systems.
This growth
induces
finer level of details,
\major{in turn inducing more complex geometrical structures in the data}.
To apprehend
this
complexity, advanced
techniques are required
\major{for the concise encoding}
of the
core
patterns
in
the data, to facilitate
analysis
\major{and}
visualization.

Topological Data Analysis (TDA) \cite{edelsbrunner09}
\major{serves this}
purpose.
It is based on robust,
multi-scale
algorithms
\cite{edelsbrunner02}, which
capture a
variety of structural features \cite{heine16}. Examples of
applications
include
combustion \cite{laney_vis06, bremer_tvcg11, gyulassy_ev14},
material sciences \cite{gyulassy_vis15, favelier16,
soler_ldav19},
 nuclear energy \cite{beiNuclear16},
fluid dynamics \cite{kasten_tvcg11, nauleau_ldav22},
bioimaging \cite{carr04, topoAngler},
data science \cite{ChazalGOS13, topoMap},
quantum chemistry \cite{chemistry_vis14, harshChemistry, Malgorzata19,
olejniczak_pccp23} \major{and}
astrophysics \cite{sousbie11, shivashankar2016felix}.

However, with the \major{above} data size increase,
it becomes
frequent in the applications that the size of a single dataset exceeds the
memory capacity of a single
computer,
hence requiring \major{the combined memories of distributed systems}.

\major{The \emph{Topology ToolKit} (TTK) \cite{ttk17} is an open-source library
which  implements a substantial collection of algorithms \cite{ttk19} for
topological data analysis and visualization. In contrast to
pre-existing, tailored, mono-algorithm implementations
(see \autoref{sec_previousWork}),
TTK
\textbf{\emph{(1)}} supports multiple algorithms (Appendix A.3),
\textbf{\emph{(2)}} it is  versatile (it provides time and memory
efficient supports for multiple, typical data representations found in scientific
computing and imaging, such as triangulated domains or regular grids)
and
\textbf{\emph{(3)}} it consistently supports  the combination of multiple
algorithms into a \emph{topological analysis pipeline} (see
the \emph{TTK Online Example Database}
\cite{ttkExamples}
for real-life examples). However, while most of its algorithms support
shared-memory parallelism \discuss{using multiple threads} with OpenMP \cite{openmp51} (Appendix
A.3), TTK did not support\major{,} up to now\major{,} distributed-memory
parallelism and thus, was
restricted to datasets of limited size, fitting in the memory of
a single computer.}

\major{This system paper addresses this issue by documenting the technical
foundations
which are required for the extension of TTK to distributed-memory parallelism
\discuss{using multiple processes} with the \emph{Message Passing Interface} (MPI), hence enabling the design of
topological pipelines for the analysis of large-scale datasets on
supercomputers.
Specifically, after formalizing our conceptual model for the distributed
representation of the input and output data (\autoref{sec_distributedSetup}), we
present the extension of TTK's internal triangulation data-structure (a central
component of its performance and versatility) to the distributed setting
(\autoref{sec_distributedTriangulation}). We also document an interface between
TTK and MPI (\autoref{sec_ttkDistributedInfrastructure}) enabling
the consistent
combination of multiple topological algorithms within a single, distributed
pipeline.}

\major{Unlike previous work (\autoref{sec_previousWork}), this paper does not
focus on the distributed computation of a specific
topological object
(such as merge trees
or persistence diagrams).
Instead, it documents the
necessary building blocks
for the extension  to the
distributed setting of a diverse
collection of
topological algorithms
such as TTK.
To evaluate
the efficiency of our extension, we document several examples
(\autoref{sec_examples}), extending to the distributed setting a selection of
topological algorithms.}
We \major{also} provide a taxonomy of TTK's topological algorithms
(\autoref{sec_taxonomy}), depending
on their communication needs and provide examples of \pierre{hybrid MPI+thread}
\pierrre{parallelizations}
for each category
(\autoref{sec_exampleImplementations}),
with detailed performance analyses (\autoref{sec_distributedAlgoPerf}). We
illustrate the new distributed
capabilities of TTK with an example of advanced analysis pipeline
(\autoref{sec_integratedPipeline}), combining
multiple algorithms, run on a dataset of \eve{120 billion vertices} distributed on
\eve{64} nodes (\autoref{sec_pipeline}) of 24 cores each. Finally, we provide a roadmap for
the completion of TTK's
MPI \julien{extension,}
with generic recommendations for each algorithm communication
category (\autoref{sec_conclusion}). This
\julien{work}
has been integrated in the main
source code of TTK and is
available in open-source.

%

\begin{figure*}
\centering
\includegraphics[width=\textwidth]{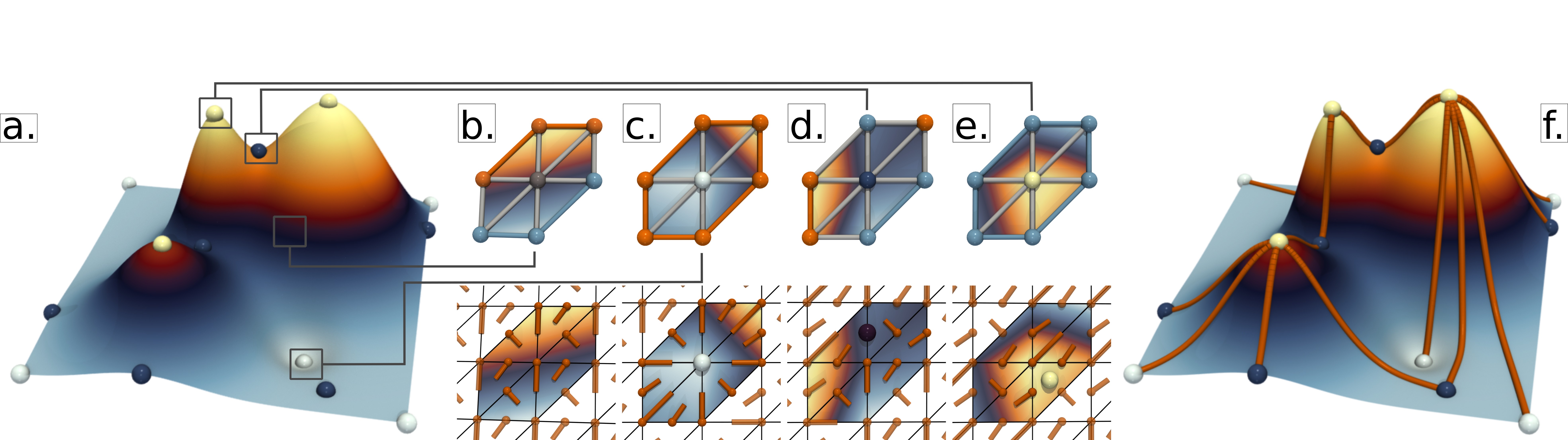}
\caption{
Topological objects considered in this paper on a
toy example (elevation $f$ on a terrain $\domain$, \emph{(a)}).
The vertices of $\domain$ can be classified
based on their star
into regular
vertices (\emph{(b)}, top: PL setting, bottom: DMT setting),
local minima \emph{(c)}, saddle points \emph{(d)} or local maxima \emph{(e)}.
Integral lines (orange curves, \emph{(f)}) are curves which are tangential to
the gradient of $f$.
}
\label{fig_toyExample}
\end{figure*}

\subsection{Related work}
\label{sec_previousWork}

Concepts
from computational topology \cite{edelsbrunner09} have
been \eve{investigated and extended} by the visualization
community~\cite{heine16} over the last two decades.
Popular topological
representations include the
persistence diagram, the Reeb graphs and its variants, or the Morse-Smale
complex \cite{edelsbrunner09}.

To improve the time efficiency of the algorithms computing the above
representations, a significant effort has been carried out to re-visit TDA
algorithms \pierre{for shared-memory parallelism}.
Several authors focused on the
shared-memory
computation of the persistence
diagram \cite{BauerKRW17, guillou_tech22}, others focused on the merge and
contour trees \major{\cite{MaadasamyDN12, AcharyaN15,
gueunet_ldav16, smirnov17,
CarrWSRFA21, gueunet_tpds19}} or
the Reeb graph \cite{gueunet_egpgv19}, while several other approaches have been
\eve{proposed for the Morse-Smale complex}
\cite{robins_pami11, ShivashankarN12, gyulassy_vis18}.
\major{Recently, a localized approach based on shared-memory parallelism   has been introduced for the on-the-fly triangulation connectivity computation \cite{LiuI24}.}
While the
above parallel approaches succeed \eve{in improving} computation times,
they
still require a shared-memory system, capable of storing the entire input
dataset \eve{into memory}. Thus, when the size of the input dataset exceeds the
capacity of the main memory \julien{of a single computer}, \pierre{distributed-memory} approaches need to be considered.
\julien{Moreover, }
\pierre{provided that the performance of these distributed
approaches scales with the
number of
nodes, they
also
\julien{contribute to reducing computation times.}}

\julien{Fewer} approaches have been documented for the computation of
topological data representations in a \pierre{distributed-memory
environment}. First, \pierre {distributed-memory}
computers are much less accessible in practice than \pierre{parallel
shared-memory architectures,}
which have become ubiquitous in recent years (workstations, laptops, etc.).
Second, the \julien{algorithmic} advances in terms of parallelism described in the
shared-memory
approaches
do not translate directly to a distributed \pierre{environment}.
Indeed, a key to the performance of the
shared-memory
approaches discussed above is the
ability of a \pierre{thread} to access any arbitrary element in the input dataset. It also
allows for easily implementable and efficient dynamic load balancing across
threads.

In contrast, in a
distributed setup,
the initial per-process decomposition of the input dataset is often
a \emph{given}, which the topological algorithm cannot modify easily and which
is likely to be unfavorable to its performances.
Then,
existing efforts for distributing TDA approaches typically consist in
first computing a \emph{local} topological representation (i.e. persistence
diagram, contour tree, etc.) given the local block of input dataset
accessible to the process and then, in a second stage, to aggregate the local
representations into a common \emph{global} representation while attempting to
minimize communications between processes
\pierre{(which are much
more costly 
than synchronizations in shared-memory
parallelism).}
Note that in several approaches \cite{MorozovW13, MorozovW14, HuangKPGBP21},
the final \emph{global} representation may not be strictly equivalent to the
output obtained by a traditional sequential algorithm, but more to a
distributed representation, capable of supporting access queries by
post-processing algorithms in a distributed fashion. Following the above
general strategy, approaches have been documented for the distributed
computation of the persistence diagram \cite{dipha} as well as the merge and
contour trees \major{\cite{PascucciC03, MorozovW13,
MorozovW14, NigmetovM19, WernerG21, HuangKPGBP21, carr22}}.
\major{However, these efforts focused on
tailored implementations (i.e. supporting a single algorithm, typically restricted to regular grids),
which neither needed to
 interact with other algorithms within a single analysis pipeline,
nor
to support compatibility with outputs computed sequentially.}
\major{For instance, DIPHA \cite{dipha} focuses on persistence diagram
computation. For that, it relies on a data representation based on the boundary
matrix of the input filtration, which
is
versatile, but at the
expense of a potentially high memory footprint. Moreover, this representation
is not accompanied by any mesh traversal functionality.
Reeber \cite{reeber, NigmetovM19} focuses on merge tree computation.
It is tailored for regular grids (with optional support of
adaptive mesh refinement
via AMReX \cite{AMReX_JOSS}) and its data structure only models vertex
adjacency relations (which is the only traversal functionality required for
merge tree computation).
In contrast, our work provides a data-structure
(\autoref{sec_distributedTriangulation}) which is
\emph{(1)} versatile (it supports both triangulated domains and regular grids),
\emph{(2)} compact and time-efficient (with an adaptive footprint for triangulated
domains and no memory overhead for regular grids),
\emph{(3)} flexible (it supports a rich set of traversals,
\autoref{ttkInitialDesign}, required to support the entire TTK algorithm collection,
Appendix A.3), and
\emph{(4)} conducive to pipeline re-use (it consistently maintains global
indices for each
simplex, irrespective of the number of processes).
}

A necessary building block for distributing TDA algorithms is an infrastructure
supporting a distributed access to the input dataset.
\major{Several general purpose software frameworks have been documented. For
instance, DIY \cite{MorozovP16} is a block-parallel library that facilitates
the parallelization of pre-existing algorithms. Specifically, DIY
enables developers to write a single implementation, which can be used for
multiple runtime configurations (out-of-core, shared-memory or
distributed-memory parallelism).
As such, DIY is a general-purpose software component, sitting right on top of
low-level parallel environments (e.g. MPI \cite{mpi40} or \discuss{C++ threads}).
In contrast to our work,
it does not provide any specific mechanism for topological algorithms.
It does not provide a distributed data-structure for simplicial complexes
(which our work contributes, \autoref{sec_distributedTriangulation}) or the
convenience functionalities needed
by arbitrary simplicial complexes
for consistently combining multiple
topological algorithms into a distributed pipeline, such as the one supported
by VTK (our work also
contributes such pipeline functionalities,
\autoref{sec_ttkDistributedInfrastructure}). Moreover,
DIY’s design philosophy eases parallelization at the cost of limiting the benefits of combining
distributed and shared-memory parallelisms. For instance, to our
understanding, workload within a single data block cannot be shared with DIY among
multiple threads (as done in contrast in our work,
\autoref{s:MPI+thread}).
To balance workload among threads with DIY, more (smaller) blocks would be
required.
This can result in
\discuss{non-optimal 
load balancing 
when the workload is not evenly distributed spatially,}
 and can be detrimental to
algorithms spanning multiple blocks (e.g. integral lines,
\discuss{\autoref{s:res_strong}}). 
}


To support topological algorithms, a
data structure must be available to efficiently traverse the input dataset,
with possibly advanced traversal queries. TTK \cite{ttk17, ttk19} implements
such a triangulation data structure, providing advanced, constant-time,
traversal queries, supporting both explicit meshes as well as the implicit
triangulation of regular grids (with no memory overhead). While several data
structures have been proposed for the distributed support of meshes
\major{\cite{stk,
IbanezSSS16, AMReX_JOSS}} (with a focus on simulation driven remeshing), we
consider in this
work the distribution of \eve{TTK's triangulation} data structure
(\autoref{sec_distributedTriangulation}), with a \julien{strong} focus on traversal time
efficiency and compatibility with a non-distributed usage, to support
post-processing interactive sessions on a workstation \julien{(c.f. \autoref{sec_distributedSetup})}.

%
%
%
%

\subsection{Contributions}
\label{sec_contributions}
This system paper makes the following new contributions.
\begin{enumerate}
 \item \emph{An efficient, distributed triangulation data structure}
(\autoref{sec_distributedTriangulation}): We introduce an extension of TTK's
\eve{triangulation data structure for the support of} distributed datasets.
 \item \emph{A software infrastructure for distributed topological
pipelines} (\autoref{sec_ttkDistributedInfrastructure}): We document a software
infrastructure
\major{consistently}
supporting advanced, distributed topological pipelines,
consisting of multiple  algorithms, possibly run on a distinct
number of processes.
 \item \emph{Examples of \pierrre{distributed} topological algorithms}
(\autoref{sec_examples}):
We provide a taxonomy of the
algorithms supported by TTK, depending
on their communication needs, and document examples of
\julien{distributed \pierrre{parallelizations,}}
with detailed
performance analyses,
\pierre{following
an MPI+thread
strategy.
This includes} an advanced pipeline consisting of multiple
algorithms, run on a dataset of \eve{120 billion vertices} on a \pierre{compute cluster} with
\eve{64} nodes
(\majorRevision{1536}
cores, total).
\item \emph{\eve{An open-source} implementation}: Our implementation
is integrated in TTK \eve{1.2.0}, to
enable others to
reproduce our results or extend TTK's distributed capabilities.
\item
\label{reproducible_example}
\eve{\emph{A reproducible example:} We provide a reference Python script of one of our advanced
pipeline\julien{s} for replicating our results with a dataset size that can be
\julien{adjusted}
to
\julien{fit the capacities of}
any system (publicly available at: \url{https://github.com/eve-le-guillou/TTK-MPI-at-example}). }
\end{enumerate}

%% file: background.tex
\section{Background}
\label{sec_background}
This section describes our formal setting and formalizes a few topological data
representations, used later in the paper when discussing examples
(\autoref{sec_examples}).
All these descriptions are given in a \emph{non-distributed} context. The
formalization of our distributed model is documented in
\autoref{sec_distributedSetup}.
We refer the reader to reference textbooks
\cite{edelsbrunner09} for a comprehensive introduction to computational
topology.


\subsection{Input data}
\label{sec_inputData}
The input is a piecewise linear (PL) scalar field $f :
\domain
\rightarrow \mathbb{R}$ defined on a $d$-dimensional simplicial
complex, with $d \leqslant 3$ in our applications
(\autoref{fig_toyExample}\emph{(a)}).
The set of $i$-simplices of $\domain$ is \majorRevision{denoted} $\domain^i$.
The \emph{star} $\Star(\simplex)$ of a simplex $\simplex$ is the set of
simplices of $\domain$ which contain $\sigma$ as a face. The \emph{link}
$\Link(\simplex)$ is the set of faces of the simplices of $\Star(\simplex)$
which do not intersect $\simplex$.
The input field $f$ is  provided on the vertices of $\domain$
and is interpolated on the simplices of higher dimension. $f$ is assumed to
be injective \julien{on the vertices}, which is achieved 
by
substituting the $f$ value
of a vertex by its position
in the
vertex order
(by increasing $f$ values).


\begin{figure*}
  \centering
   \includegraphics[width=\textwidth]{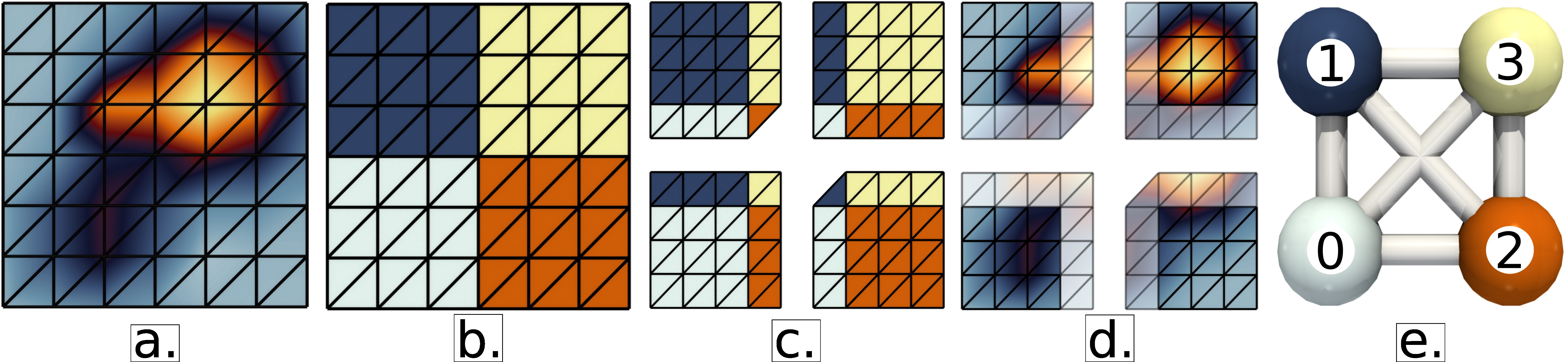}
  \caption{
The input data \emph{(a)} is assumed to be loaded in
the memory of $n_p$ independent processes in the form of $n_p$ disjoint
\emph{blocks} of data (\emph{(b)}, one color per block, $n_p = 4$ in this
example). A layer of \emph{ghost} simplices (\emph{(c)}, coming from
adjacent blocks, matching colors) is added to each block.
This local data duplication \eve{(\emph{(d)}, transparent)} eases subsequent
processing
on
block boundaries\eve{. A} local adjacency
graph is constructed to encode
local
neighbor relations between blocks \emph{(e)}.
 }
 \label{fig_distributedModel}
 \end{figure*}

\subsection{Critical points}
\label{sec_criticalPoints}
The sub-level set
$\sublevelset{f}(w)$
of an isovalue $w \in \mathbb{R}$ is
defined as $\sublevelset{f}(w) = \{p \in \mathcal{M} ~|~ f(p) < w\}$. It can be
interpreted as a subset of the data, below the isovalue $w$.
As $w$ continuously increases, the topology of $\sublevelset{f}(w)$ changes at
specific vertices of $\domain$, called the \emph{critical points} of $f$.
Let $\Link^-(v)$ be the \emph{lower link} of the vertex $v$:
$\Link^-(v) = \{\simplex \in \Link(v) ~|~ \forall u \in
\simplex : f(u) < f(v)\}$ (blue edges and vertices in
\autoref{fig_toyExample}\emph{(b-e)}, top).
The \emph{upper link} of $v$ is defined symmetrically:
$\Link^+(v) = \{\simplex \in \Link(v) ~|~ \forall u \in
\simplex : f(u) > f(v)\}$ (orange edges and vertices in
\autoref{fig_toyExample}\emph{(b-e)}, top).
A vertex $v$ is \emph{regular} if and only if both $\Link^-(v)$ and
$\Link^+(v)$ are simply connected. Otherwise, $v$ is a
\emph{critical vertex} of
$f$ \cite{banchoff70}.
A critical vertex $v$ can be classified by its \emph{index} $\Index(v)$, which
is
$0$ for minima (\autoref{fig_toyExample}\emph{(c)}), $1$ for $1$-saddles
(\autoref{fig_toyExample}\emph{(d)}), $(d-1)$ for $(d-1)$-saddles and $d$
for maxima (\autoref{fig_toyExample}\emph{(e)}). Vertices for which the number
of connected components of
$\Link^-(v)$ or $\Link^+(v)$ are greater than $2$ are called \emph{degenerate
saddles}.
\major{Prior to this work, this critical point classification 
was implemented in TTK with shared-memory parallelism with OpenMP (see 
Appendix A.3), as each vertex classification is independent.}

%

\subsection{Integral lines}
\label{sec_integralLines_def}
Integral lines are curves on $\domain$ which
locally describe the gradient of $f$ (orange curves in
\autoref{fig_toyExample}\emph{(f)}).
They can be used
to capture and visualize
adjacency relations between critical points.
\eve{The starting vertex of an integral line is called a \emph{seed}.}
Given a \julien{seed} $v$, its \emph{forward} integral line, noted
$\forwardIntegralLine(v)$, is a path along the edges of $\domain$, initiated in
$v$, such that each edge of $\forwardIntegralLine(v)$ connects a vertex $v'$ to
its highest neighbor $v''$.
When encountering a saddle $s$, we say that an integral line \emph{forks}: it
yields one new integral line per connected component of $\Link^+(s)$.
Integral lines can \emph{merge} (and possibly fork later).
A \emph{backward} integral line, noted
$\backwardIntegralLine(v)$, is defined symmetrically (i.e. integrating
downwards).
\major{Prior to this work, this computation
was implemented sequentially in TTK (see Appendix A.3), i.e. 
given a list of input seeds,
each integral line 
was computed sequentially one after the other.}


\subsection{Discrete gradient}
\label{sec_discreteGradient}
In recent years, an alternative emerged to the PL formalism of critical points
described above (\autoref{sec_criticalPoints}), namely Discrete Morse Theory
(DMT)
\cite{forman98}.
This formalism implicitly resolves several challenging configurations (such as
degenerate saddles on manifold domains), which has been particularly useful for
the development of robust algorithms in the context of Morse-Smale complex
computation \cite{robins_pami11, gyulassy_vis18}. We also consider in
this work this alternative representation to critical points, as it nicely
exemplifies a large set of the traversal features supported by TTK's
triangulation (\autoref{sec_distributedTriangulation}).

A \emph{discrete vector} (small orange arrows,
\autoref{fig_toyExample}\emph{(b-e)}, bottom) is a pair formed by a simplex
$\simplex_i \in
\domain$ (of dimension $i$) and one of its co-facets $\simplex_{i+1}$ (i.e. one
of its co-faces of dimension $i+1$), noted $\{\simplex_i < \simplex_{i+1}\}$.
$\simplex_{i+1}$ is usually referred to as the \emph{head} of the vector
(represented with a small orange cylinder in
\autoref{fig_toyExample}\emph{(b-e)}, bottom), while
$\simplex_{i}$ is its \emph{tail} (represented with a small orange sphere in
\autoref{fig_toyExample}\emph{(b-e)}, bottom). Examples of discrete vectors
include a pair
between a vertex and one \pierre{of} its incident edges, or a pair between an edge and a
triangle containing it. A \emph{discrete vector field} on $\domain$ is then
defined as a collection $\discreteVectorField$ of pairs $\{\simplex_i <
\simplex_{i+1}\}$, such that each simplex of $\domain$ is involved in at most
one pair. A simplex $\simplex_i$ which is involved in no discrete vector
$\discreteVectorField$ is called a \emph{critical simplex}.
A
\emph{v-path} is a sequence of discrete vectors
$\big\{\{\simplex^0_i < \simplex^0_{i+1}\}, \dots,
\{\simplex^k_i <
\simplex^k_{i+1}\}\big\}$, such that \emph{(i)} $\sigma^j_i \neq
\sigma^{j+1}_i$ (i.e. the tails of two consecutive vectors are distinct) and
\emph{(ii)} $\sigma^{j+1}_i < \sigma^{j}_{i+1}$ (i.e. the tail of a vector in
the sequence is a face of the head of the previous vector), for any $0 < j <
k$. A \emph{discrete gradient field} is
a discrete vector field such that
all its possible \emph{v-paths} are loop-free. Several algorithms have been
proposed
to compute such a discrete gradient field from an
input PL scalar field. We consider in this work the algorithm by Robins et al.
\cite{robins_pami11}, given its proximity to the PL setting: each
critical cell identified by this algorithm is guaranteed to be located in the
star of a PL critical vertex (\autoref{sec_criticalPoints}).
\major{Prior to this work, this computation
was implemented in TTK with shared-memory parallelism with OpenMP 
(Appendix A.3), as each lower star can be processed 
independently.}

%% file: setup.tex
\section{Distributed Model}
\label{sec_distributedSetup}

We now formalize our distributed model, which will eventually be used
\julien{as a blueprint}
to port
%
%
the algorithms
described above
(\autoref{sec_background}) to distributed computations
(\autoref{sec_examples}).

\subsection{Input distribution formalization}
\label{sec_distributedInputFormalization}


\subsubsection{Decomposition}

\major{Our distributed-memory model is based the following convention.}
$f$ \eve{is \julien{assumed to be} loaded} in the memory of
$\numberProcesses$
processes in the form of $\numberProcesses$
disjoints \emph{blocks} of data (\autoref{fig_distributedModel}\emph{(a-b)}).
Specifically, each process $i \in \{0, \dots,
\numberProcesses-1\}$ is associated with a local block $f_i : \domain_i
\rightarrow \mathbb{R}$,
such that:
\begin{itemize}
 \item $\domain_i \subset \domain$: each block \major{$\domain_i$} is a
 $d$-dimensional simplicial complex,
 being a subset of the global input \major{$\domain$}.
 \item  \major{Any simplex $\simplex$ present in multiple blocks (e.g. at the boundary between adjacent blocks) is said to be \emph{exclusively owned}, by convention, by the process with the lowest identifier (among the processes containing $\simplex$).
  \item A simplex $\simplex \in \domain_i$ which is not exclusively owned by  the process $i$ is called a \emph{ghost simplex} (\autoref{sec_ghostedBlock}).}
  \item $\cup_{\domain_i} = \domain$: the union \eve{of the blocks} is equal to
the input.


\end{itemize}


\subsubsection{Ghost layer}
\label{sec_ghostedBlock}

\majorRevision{In such a distributed setting, \emph{ghost} simplices are typically considered,
in order to save communications
between processes for local tasks.
Ghost simplices are typically simplices inside the block of a process that are copies
of the interfacing simplices of an adjacent block (see the lighter simplices in \autoref{fig_distributedModel}, (d)). }
We note $\domain'_i$ the
$d$-dimensional simplicial complex obtained by considering a layer of
ghost simplices, i.e.
by adding to $\domain_i$
\major{the}
$d$-simplices \major{of $\domain$ which share a face with a $d$-simplex of $\domain_i$,}
along with all their
$d'$-dimensional faces (with $d' \in \{0, \dots, d-1\}$).
Overall, all the simplices added in this way to the block $\domain_i$ to form
the \emph{ghosted block} $\domain_i'$ are  \emph{ghost simplices}
(\autoref{fig_distributedModel}\emph{(c-d)}).

The usage of such a
ghost layer is typically motivated in practice by algorithms which perform
local traversals (e.g. PL critical point extraction,
\autoref{sec_criticalPoints}). Then, when such algorithms reach the boundary of
a block, they can still perform their task without any communication, thanks
to the ghost layer.
Also, the usage of a ghost layer facilitates the
identification of boundary simplices (i.e. located on the boundary of
the \emph{global} domain $\domain$, see \autoref{sec_explicitPreconditioning}).

\eve{The blocks are also positioned in relation to one another.
Processes $i$ and $j$ will
be considered adjacent (\autoref{fig_distributedModel}\emph{(e)}) if $\domain_i'$ contains $d$-simplices that are
exclusively owned by $j$
and if $\domain_j'$ contains
$d$-simplices that are
exclusively owned by $i$.}







\subsubsection{Global simplex identifiers}
\label{sec_globalIdentifiers_spec}

For any $d' \in \{0, \dots, d\}$, each $d'$-simplex $\sigma_j$ of
each block
$\domain'_i$ is associated with a \emph{local} identifier $j \in [0,
|\domain{'}_i^{d'}|-1]$.
This integer uniquely identifies $\sigma_j$ within the local block $\domain'_i$.


The simplex $\sigma_j$ is also associated with a \emph{global} identifier
$\phi_{d'}(j) \in [0, |\domain^{d'}|-1]$,  which uniquely identifies $\sigma_j$
within the \emph{global} dataset $\domain$. Such a global identification is
motivated by the need to support varying numbers of processes.
In particular, assume that a first analysis pipeline $P_1$ (for instance
extracting critical vertices, \autoref{sec_criticalPoints}) uses $n_p(P_1)$
processes to generate an output (e.g. the list of critical vertices).
Let us consider now a second analysis pipeline $P_2$  using $n_p(P_2)$
processes (possibly on a different machine) to post-process the output of
$P_1$ (for instance, seeding
integral lines, \autoref{sec_integralLines}, at the previously extracted
critical vertices). Since \eve{$n_p(P_1)$} and $n_p(P_2)$ differ between the two sub-pipelines,
their
input decompositions into local blocks will also differ. Then the local
identifiers of the \eve{critical vertices} employed in $P_1$ may no longer be usable in $P_2$.
For instance, if $n_p(P_1) < n_p(P_2)$, the local blocks of $P_2$ may be much
smaller than those of $P_1$ and the local identifiers of $P_1$ \julien{can} become out
of range in $P_2$. Thus, a common ground between the two pipelines need to be
found to reliably exchange \eve{information, hence} the global,  unique identifiers.

Note that the support for a
varying number of processes is a necessary feature for
\pierre{practical distributed}
topological algorithms. While it is a challenging
constraint (c.f. \autoref{sec_distributedTriangulation}), it
is
beneficial to
%
various application use cases. For instance,
$P_2$ can be a post-processing pipeline run on a
\eve{workstation.} $P_2$ can
also
be executed on a different (possibly larger)
\pierre{distributed-memory} system
than $P_1$. Last,
$P_1$ and
$P_2$ \eve{can be} part of a single, large pipeline, which would
include an aggregation step of the outputs of $P_1$ to a different number of
processes ($n_p(P_2)$).



\subsubsection{\major{Simplex-to-process maps}}

Each block $\domain'_i$ is associated with
\majorRevision{\emph{simplex-to-process maps},
which map each simplex} to the identifier of the process which \major{exclusively} owns it.




\subsection{Output distribution formalization}
\label{sec_distributedOutput}
Topological algorithms typically consume an input (possibly complex), to
produce a (usually) simpler output (such as the topological
representations described in \autoref{sec_background}). Moreover, multiple
topological algorithms can be combined sequentially to form an analysis
pipeline. For instance, a first algorithm $A_1$ may compute integral lines
(\autoref{sec_integralLines}) for a first field $f$, while a second algorithm
$A_2$ may extract the critical vertices (\autoref{sec_criticalPoints}) for a
second field $g$, defined on the integral lines generated by the first
algorithm $A_1$. Thus, the output produced by a distributed topological
algorithm $A_1$ \emph{must} be
readily
usable by another distributed algorithm $A_2$.

This implies that
the output computed by a topological algorithm must also strictly comply
to the input specification \eve{(\autoref{sec_distributedInputFormalization}) and
should contain:} \emph{(i)} a ghost layer,
\emph{(ii)}
global simplex identifiers,
and \emph{(iii)}
\majorRevision{simplex-to-process maps}.

Note that, according to this formalism, the output of a topological algorithm
\emph{is} distributed among several processes. Depending on the complexity of
this output, specialized manipulation algorithms (handling communication
between processes) may need to be later developed to exploit them appropriately
in a post-process.

\subsection{Implementation specification}
\label{sec_implementationSpecification}
We now review the building blocks which are necessary
to support
the distributed model
specified in Secs. \ref{sec_distributedInputFormalization} and
\ref{sec_distributedOutput}.

\eve{The pipeline combining the different topological
algorithms can be encoded in the form of a Python script \julien{(c.f. contribution \ref{reproducible_example}, \autoref{sec_contributions})}.}
The initial decomposition of the global domain $\domain$ and the ghost layer
(specifically, the ghost vertices and the ghost $d$-simplices)
are computed by ParaView \cite{paraviewBook}. \eve{\julien{Then,} the TTK algorithms present in the pipeline will be
instantiated by ParaView on each process and from this point on, they will
be able to access their own local  block of \emph{ghosted} data and
communicate with other processes. }

While ParaView offers in principle the possibility to compute
vertex-to-process maps,
we have observed several inconsistencies (in particular
when
using ghost layers), which prevented us to use it reliably.  This
required us to develop our own process identification strategy
(\autoref{sec_rankArrays}).

Moreover, while ParaView also offers in principle the possibility to generate
global identifiers for vertices and cells (i.e. $d$-simplices), we
have experienced technical difficulties with it (such as
a dependence of the resulting identifiers on the number of processes), as well
as issues which made it unusable for large-scale datasets (such as an
excessively large memory footprint). This required us to develop our own
strategy for the global identification of vertices and cells (i.e.
$d$-simplices), documented in
Secs. \ref{explicitTriangulation} and \ref{implicitTriangulation}.

The input PL scalar field $f$ is
required to be injective on the vertices (c.f. \autoref{sec_inputData}).
This can be
\major{easily obtained via lexicographic vertex comparison,
by considering for each vertex $v$ the tuple
$\big(f(v), \phi_0(v)\big)$, i.e. the tuple
formed by its scalar value and its global identifier.
In practice, to accelerate these comparisons for \emph{local} vertices
(i.e. vertices
present in a common block $\domain'_i$),
the process $i$ will
%
first sort all its local vertices (in lexicographic order) in a preconditioning step, and
local vertex comparisons will later be based
on their order in the sorted  list.}



\autoref{sec_distributedTriangulation}
documents the extension of TTK's triangulation data structure to support our
model of distributed input and output (Secs.
\ref{sec_distributedInputFormalization} and \ref{sec_distributedOutput}).

\major{Additional procedures
easing the combination of
multiple algorithms into a single pipeline 
(adjacency graph computation, ghost data exchange) are documented in 
\autoref{sec_lowLevelInterface}.}

%% file: triangulation.tex
\begin{figure*}
\includegraphics[width=\linewidth]{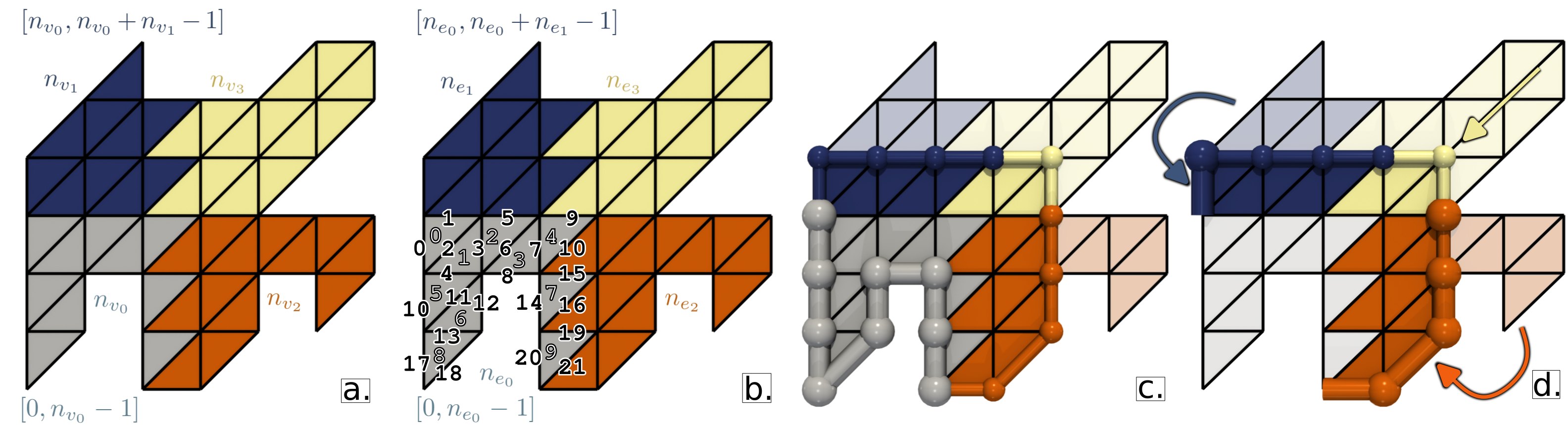}
\caption{Preconditioning of our distributed explicit triangulation.
\emph{(a)}
Each process $i$ enumerates its number \eve{$n_{v_i}$} \pierre{of}
\julien{exclusively}
\eve{owned} vertices and
$d$-simplices. Next, an MPI prefix sum provides a local offset for each process
to generate global identifiers. \emph{(b)} For each process $i$,
simplices of intermediate dimensions (edges \eve{($n_{e_i}$)}, triangles) are 
locally enumerated
for contiguous intervals of global identifiers of $d$-simplices \eve{(white 
numbers)}.
Next, all the intervals are sent to
the process $0$ which sorts them \pierre{first} by 
\majorRevision{simplex-to-process} identifier, then by interval
start, yielding a per-interval offset that each process can use to generate its
global identifiers \eve{(black numbers)}. \emph{(c)} Within a given block, the 
vertices at the
boundary of the domain \eve{$\domain$} are identified as non-ghost boundary 
vertices \eve{(large spheres)}. 
Next, a simplex which only contains boundary vertices is considered to be a 
boundary
simplex \eve{(larger cylinders)}. \emph{(d)} The global identifiers and boundary 
information of the
ghost simplices are retrieved through MPI communications with the neighbor 
\pierre{processes}.
\eve{The
\julien{ghost}
simplices on the global boundary are flagged as boundary simplices (larger 
spheres and cylinders)}.}
\label{fig_explicitTriangulation}
\end{figure*}

\section{Distributed Triangulation}
\label{sec_distributedTriangulation}

This section describes the distributed
extension
of TTK's triangulation
data structure, later used by each topological algorithm.
In the following, we assume that the input block is loaded in the memory of the
local process $i$ and ghosted (i.e. we consider the ghosted block
$\domain'_i$, \autoref{sec_ghostedBlock}).
\julien{Moreover,
we consider that, for each process $i$, a list of \emph{neighbor processes} is
available
(\autoref{fig_distributedModel}\emph{(e)}).}

\subsection{Initial design}
\label{ttkInitialDesign}
For completeness, we briefly summarize
the 
\major{pre-existing}
implementation of TTK's triangulation data structure \major{\cite{ttk17}.} 
\major{In the following, the triangulation $\domain$ is assumed to be of uniform top dimension, i.e. any $d'$-simplex (with $d' \in \{0, 1, \dots, d-1\}$) admits at least one $d$-dimensional co-face.}

In the explicit case (the input is a simplicial mesh),
this data structure takes as an input a pointer to an array of 3D points
(modeling the vertices of $\domain$), as well as a pointer to an array of
indices (modeling the $d$-simplices of $\domain$).
In the implicit case (the input is a regular grid), it takes as an input the
origin of the grid as well as its resolution and spacing across each dimension.
These can be provided by any
IO library (in our experiments, these are provided by VTK).

Based on this input, the triangulation supports a variety of traversal
routines, to address the needs of the \eve{algorithms}.
\begin{enumerate}
  \item \textbf{Simplex enumeration:} for any $d' \in \{0, \dots, d\}$, the
data structure can enumerate all the $d'$-simplices of $\domain$.
  \item \textbf{Stars and links:} for any $d' \in \{0, \dots, d\}$, the data
structure can enumerate all the simplices of the star and the link of any
$d'$-simplex $\simplex$.
  \item \textbf{Face / co-face:} for any $d' \in \{0, \dots, d\}$, the data
structure can enumerate all the $d''$-simplices $\tau$ which are faces or
co-faces of a $d'$-simplex $\simplex$, for any dimension $d''$ (i.e. $d'' \neq
d'$ and $d'' \in \{0, \dots, d\}$).
  \item \textbf{Boundary tests:} $d' \in \{0, \dots, d - 1\}$, the
data structure can be queried to determine if a $d'$-simplex $\simplex$ is on
the boundary of $\domain$ or not.
\end{enumerate}

As discussed in the original
paper \cite{ttk17}, such traversals are
rather typical of topological algorithms, which may need to inspect
extensively the local neighborhoods of simplices.

All
traversal queries (e.g. getting the $i^{th}$ $d''$-dimensional co-face of a
given $d'$-simplex $\simplex$) are addressed by the data structure in constant
time, which is of paramount importance to guarantee the runtime performance of
the calling topological algorithms. This is supported by the data structure via
a \major{\emph{preconditioning} mechanism} \major{(i.e. an adaptive, pre-computation stage,
described in Appendix A.2). For regular grids, periodic (along all dimensions) or not,
the traversal queries can be computed on-the-fly given the regular structure of 
the Freudenthal
triangulation of the grid \cite{freudenthal42, kuhn60}.}

\subsection{Distributed explicit triangulation}
\label{explicitTriangulation}
This section describes our distributed implementation of the TTK triangulation
in explicit mode, i.e. when an explicit simplicial complex is provided as a
global input.

\subsubsection{\julien{Distributed explicit preconditioning}}
\label{sec_explicitPreconditioning}
The preconditioning of explicit triangulations in the distributed setting
involves the computation of four main \majorRevision{pieces of information}: \emph{\textbf{(1)}} global
identifiers,
\emph{\textbf{(2)}} ghost global identifiers,
\emph{\textbf{(3)}} boundary,
and \emph{\textbf{(4)}} ghost boundary.

\noindent
\textbf{\emph{(1)} Global identifiers:}
The first step
consists in
determining global identifiers for the vertices (i.e.,
the map $\phi_{0}$,
\autoref{sec_distributedInputFormalization},
\major{}
its inverse,
$\phi^{-1}_0$). This step is not optional and is triggered automatically.
\major{For} each
ghosted block $\domain_i'$, the number
\eve{$n_{v_i}$} of \emph{non-ghost} vertices
that the block exclusively owns is computed \eve{(\autoref{fig_explicitTriangulation}\emph{(a)})}.
\julien{Next, an MPI
prefix sum is performed to
determine the offset
that each block $i$ should add to its local
vertex
identifiers to obtain its global vertex identifiers.} 
The map $\phi_d$ and its inverse
$\phi^{-1}_d$ are computed similarly.

\noindent
\julien{Next, global identifiers need to be computed for the 
\major{$d'$-}simplices \major{of intermediate dimension (i.e. $d' \in
\{1, \dots, d-1\}$, \autoref{fig_explicitTriangulation}\emph{(b)}))}.}
%
This step is optional and
is only triggered if the calling algorithm pre-declared the usage of these
simplices in the preconditioning phase. 

\major{For this, each process $i$ first identifies,}
among its list of \julien{exclusively} \eve{owned} $d$-simplices,
intervals of contiguous global identifiers. These are
typically interleaved with global identifiers of ghost $d$-simplices.
\major{Then, intervals are processed independently via shared-memory parallelism, and for}
each
interval $x$, the $d'$-simplices are
provided with a local
identifier (with the same procedure as used in the non-distributed setting).
Given a $d'$-simplex $\simplex$ at the interface between two blocks (i.e.
$\simplex$ is a face
\pierre{of} a ghost $d$-simplex), a tie break strategy needs
to be established, to guarantee that only one process tries to generate an
identifier for $\simplex$.
Specifically,
the process $i$ will generate an
identifier for $\simplex$ only if $i$ is the lowest 
\majorRevision{simplex-to-process} identifier among
the exclusive owners of the $d$-simplices in $\Star(\simplex)$ \major{(\autoref{sec_distributedInputFormalization})}.
Next, all the intervals (along with their \majorRevision{simplex-to-process} 
identifier and number of
$d'$-simplices) are sent to
the process $0$ which, after ordering the intervals of $d$-simplices
\pierre{first} by
\majorRevision{simplex-to-process} identifier then by local identifier, 
determines
the offset
that each interval $x$ should add to its local
$d'$-simplex identifiers to obtain its global identifiers.

\noindent
\textbf{\emph{(2)} Ghost global identifiers:}
The second step of the preconditioning consists in retrieving for a given block
$\domain_i'$ the global identifiers of its ghost 
\major{$0-$ and $d-$}simplices. \eve{This step is not
optional and is always triggered.} 
This
feature can be particularly useful when performing
local computations on the boundary of the block (e.g. discrete gradient,
\autoref{sec_examples}).

\noindent
Once all the processes have established their vertex global identifiers, each
process $i$
queries each of its neighbor processes $j$, to obtain the global identifiers of
its ghost vertices \major{(a KD-tree data-structure is employed to establish,
with shared-memory parallelism,
the correspondence between vertices coming from different blocks)}.
Once global vertex identifiers are available for the ghost vertices of
$\domain'_i$, a simpler 
exchange 
procedure 
\major{is used}
to collect the
global identifiers of 
\major{the
ghost $d'$-simplices with $d' \in
\{1, \dots, d\}$ (the correspondence between $d'$-simplices coming from 
different blocks is established,
with shared-memory parallelism,
based on the global identifiers of their
vertices).}


\noindent
\textbf{\emph{(3)} Boundary:}
The third  step consists in determining the simplices which are on
the boundary
of the global domain $\domain$. This step is optional and is only triggered (on
a per simplex dimension basis) if
the calling algorithm pre-declared the usage of boundary simplices in the
preconditioning phase.
This feature is particularly useful for algorithms which process as
special cases
the simplices which are \eve{on the boundary
of $\domain$} (e.g. critical point extraction,
\autoref{sec_examples}).

\noindent
\major{Each}
process $i$ identifies the boundary vertices of its \emph{ghosted block}
$\domain_i'$
\eve{(See \autoref{fig_explicitTriangulation}\emph{(c)})}, with \majorRevision{exactly the same procedure as the one} used
in
the non-distributed setting \cite{ttk17}.
Then, thanks to the ghost layer, it is guaranteed that among the 
set of
\major{boundary}
vertices
\major{identified above},
the non-ghost vertices are indeed on the boundary of the global
domain $\domain$.
Finally, 
\major{a}
$d'$-simplex
will be
marked as a boundary simplex if all its vertices are on the boundary of
$\domain$.

\noindent
\toDelete{\textbf{\emph{(4)} Ghost boundary:}
Similarly to step \emph{(2)},
a final step of data exchange between the process $i$ and its neighbors
enables the retrieval of the ghost simplices of $\domain_i'$ which are also on
the boundary \eve{(\autoref{fig_explicitTriangulation}\emph{(d)})}. This step is optional and is only triggered if
the calling algorithm pre-declared the usage of boundary 
simplices in the
preconditioning phase.}

Finally, the preconditioning of any other traversal routine is identical to the
non-distributed setting.

\subsubsection{Distributed explicit queries}
\label{sec_explicitQueries}
In this section, we describe the implementation of the traversals of the
triangulation, as queried by a calling algorithm. This assumes that the calling
algorithm first called the appropriate preconditioning functions in a
pre-process.

The traversal of a local ghosted  block $\domain'_i$ by an algorithm
instantiated on the process $i$ is performed
identically to the
non-distributed setting, with local simplex identifiers.
\major{This requires the calling algorithm to locally translate input (and output)
\emph{global} simplex identifiers into \emph{local} ones (i.e. with the maps 
introduced in \autoref{sec_globalIdentifiers_spec})}.


\subsection{Distributed implicit triangulation}
\label{implicitTriangulation}
This section describes our distributed implementation of the TTK triangulation
in implicit mode, i.e. when a regular grid is provided as a global input. Then,
as described below,
most traversal information
can be \majorRevision{computed on-the-fly} at runtime,
given
the regular sampling pattern of 
\major{the}
Freudenthal
triangulation \cite{freudenthal42, kuhn60} of the input grid.

\begin{figure}
\includegraphics[width=\linewidth]{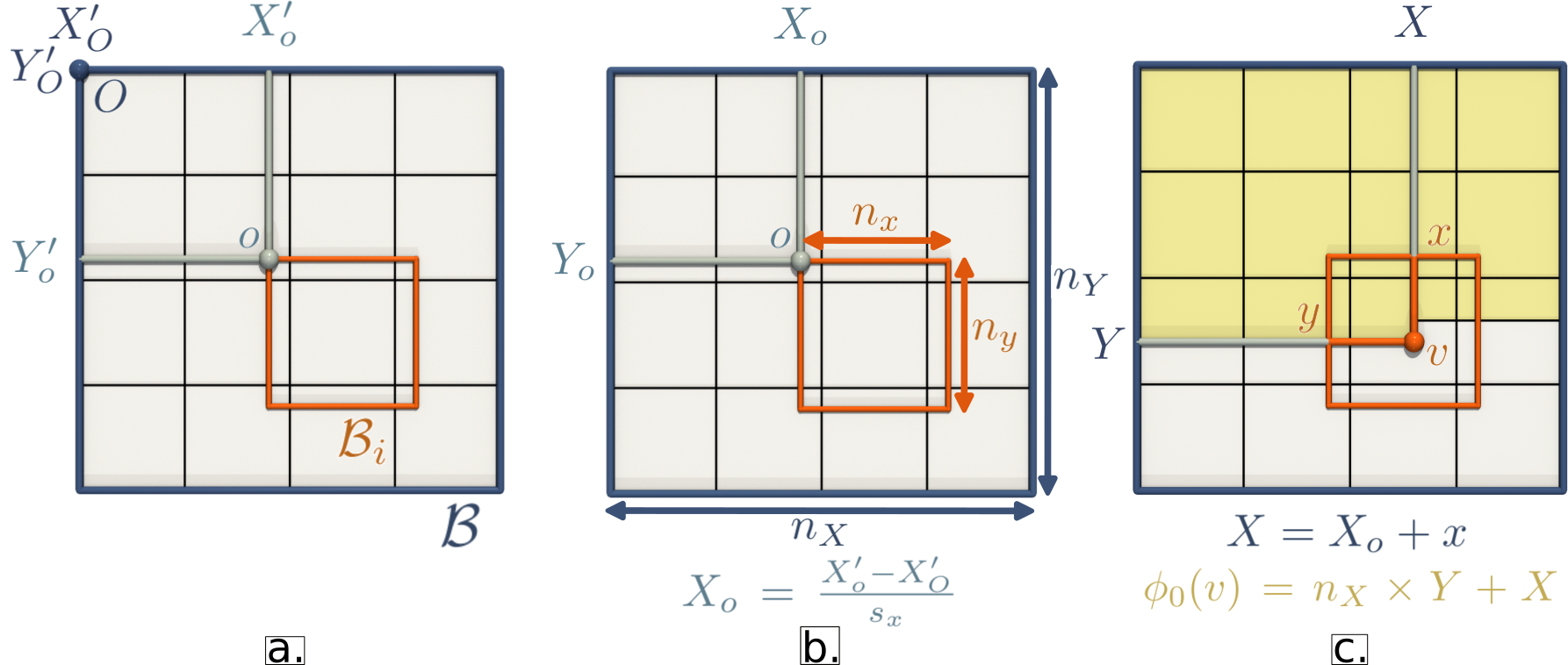}
\caption{\eve{Preconditioning of our distributed implicit triangulation. \emph{(a)}
Each process $i$ computes
\major{(with shared-memory parallelism)}
the bounding box $\boundingBox_i$ of its ghosted
block $\domain_i'$.  The vertex $o$, respectively $O$, is the origin of $\domain_i'$, respectively $\domain$,
 with $(X'_o, Y'_o, Z'_o)$, respectively $(X'_O, Y'_O, Z'_O)$, its floating-point coordinates. 
 The bounding box $\boundingBox$ of $\domain$ is computed \major{(via MPI 
parallel reductions)} from all the local $\boundingBox_i$.
\emph{(b)} Two key pieces of information are computed at this step: the dimensions of the global grid $(n_X, n_Y, n_Z)$
(the number of vertices of $\domain$ \eve{in each direction})
and the local grid offset $(X_o, Y_o, Z_o)$ (the global discrete coordinates of $o$). It is computed from $(X'_O, Y'_O, Z'_O)$, $(X'_o, Y'_o, Z'_o)$ and
the floating-point spacing of the grid $(s_x, s_y, s_z)$.
Following that, each process locally instantiates a global implicit triangulation model
of $\domain$.
\emph{(c)} Given a local vertex identifier, its global discrete coordinates $(X, Y, Z)$ in 
$\domain$ are inferred from its local discrete point coordinates $(x, y, z)$ (with $x \in [0, n_x -1]$, $y \in [0, n_y -1]$, and $z \in [0, n_z -1]$,
$n_x$, $n_y$ and $n_z$ being the number of vertices of the grid $\domain_i'$ in each direction), and its \eve{local grid} offsets.
Next, its global identifier, \eve{$\phi_0(v)$}, is determined on-the-fly by global row-major
indexing.}}
\label{fig_implicitTriangulation}
\end{figure}

\subsubsection{Distributed implicit preconditioning}
\label{sec_implicitPreconditioning}
In implicit mode, the preconditioning of the triangulation 
\major{identifies}
the position of the local ghosted grid
$\domain'_i$ within the global grid $\domain$\major{, as detailed in 
\autoref{fig_implicitTriangulation}}. This step is not optional and is
triggered automatically.
The preconditioning of any traversal routine returns immediately
without any processing (all queries are \majorRevision{computed on-the-fly}).

\subsubsection{Distributed implicit queries}
In this section, we describe the implementation of the traversals of the
triangulation, as queried by a calling algorithm.

The traversal of a local ghosted  block $\domain'_i$ by an algorithm
instantiated on the process $i$ is performed
identically to the
non-distributed setting, with local simplex identifiers.

Similarly to the explicit case (\autoref{sec_explicitQueries}), 
\major{the calling algorithm must now translate input (and output) 
\emph{global} simplex identifiers into \emph{local} ones (i.e. with the maps 
from
\autoref{sec_globalIdentifiers_spec})}.


The important difference with the explicit mode is that all the information
computed in explicit preconditioning
(i.e. \textbf{\emph{(1)}} global identifiers, \textbf{\emph{(2)}} ghost global
identifiers, \textbf{\emph{(3)}} boundary, and \textbf{\emph{(4)}} ghost
boundary,  see \autoref{sec_explicitPreconditioning}) now needs to be \majorRevision{computed on-the-fly} 
at runtime (i.e. upon the query of this information by the calling algorithm).

\noindent
\textbf{\emph{(1)} Global identifiers:} \eve{\major{As detailed in 
\autoref{fig_implicitTriangulation}, given}
a local vertex $v$, its global discrete coordinates $(X, Y, Z)$ 
in the global
grid $\domain$ are inferred from its local discrete point coordinates $(x, y, 
z)$ \major{in $\domain_i'$ (\autoref{fig_implicitTriangulation}\emph{(c)}),}
and the local grid offset 
$(X_o, Y_o, Z_o)$. 
\major{From the coordinates} $(X, Y, Z)$, the global identifier of $v$ is
computed on-the-fly with the 
procedure used in the non-distributed setting
\cite{ttk17} (global row-major indexing).
The same \major{procedure is used} 
for $d$-simplices.}

\noindent
The global identifier of any 
$d'$-simplex
\eve{($d' \in \{1, \dots, d-1\}$)} is computed
\major{by identifying the $d'$-simplex in $\domain$  which has the same
global vertex identifiers
(via vertex star inspection).}

\noindent
\textbf{\emph{(2)} Ghost global identifiers:} The global
identifier of a ghost \major{simplex is also computed with the above procedure.}


\noindent
\textbf{\emph{(3)} Boundary:} To decide if a given $d'$-simplex
is on the boundary of $\domain$, its global identifier is first
retrieved (above) and the local copy of the global grid
$\domain$ is queried for boundary check based on this global identifier (with
the exact procedure used in the non-distributed setting \cite{ttk17}).

\noindent
\toDelete{\textbf{\emph{(4)} Ghost boundary:} The boundary check for ghost simplices is
\major{also} computed with the \major{above procedure.}
}

\begin{figure}
  \includegraphics[width=\linewidth]{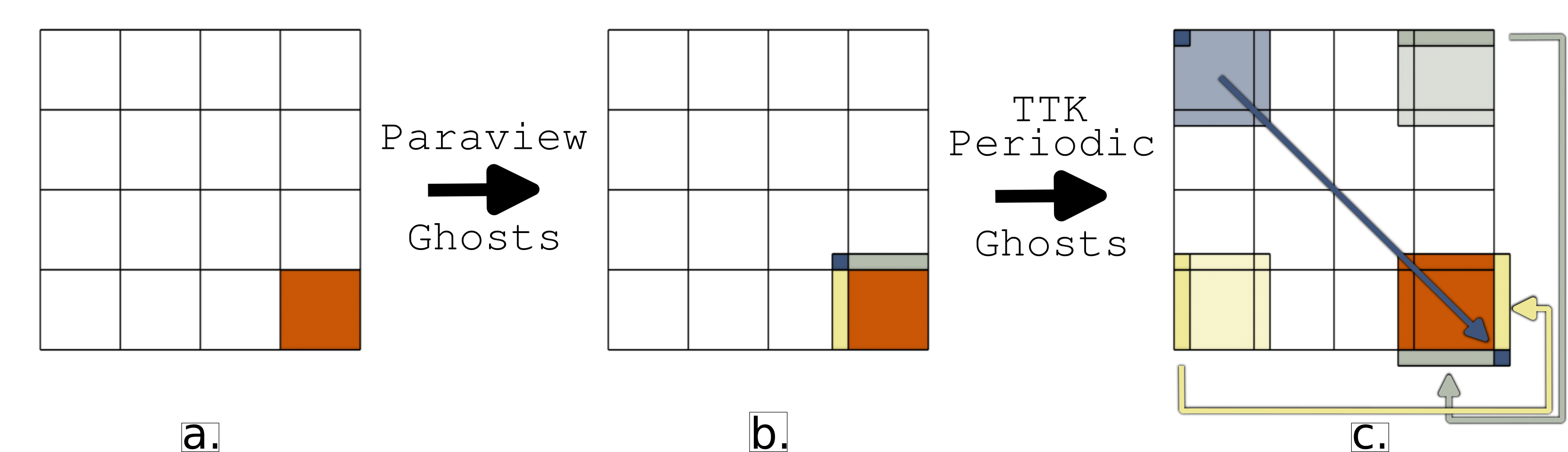}
 \caption{Preconditioning of our distributed periodic implicit triangulation.
This triangulation type is handled similarly to the implicit case, but
additional ghost simplices need to be computed.
Given a data block
$\domain_i$ \eve{(\emph{(a)}, orange)}, ParaView generates a first layer of ghost
$d$-simplices (\emph{(b)}, blue, grey, yellow). If $\domain_i$ was located on the boundary of the
global grid $\domain$, periodic boundary conditions must be considered by
adding an extra layer of ghost $d$-simplices \eve{(arrows)} for each periodic face
of $\domain$ \emph{(c)}.}
 \label{fig_periodicCondition}
\end{figure}

\subsection{Distributed implicit periodic triangulation}
\major{Periodic grids (with periodicity in all dimensions) are supported via implicit Freudenthal triangulation
\cite{freudenthal42, kuhn60} like in the previous section. However, the
periodic boundaries require
specific
adjustments in terms of
preconditioning.}


\major{Since ParaView's ghost cell generator only produces ghosts at the 
interface of 
the domain of processes, an extra layer of ghost simplices needs to be computed,} 
as illustrated in \autoref{fig_periodicCondition}.
Specifically, each process $i$ checks if its block $\domain_i'$ is located on
the boundary of the global grid $\domain$ \major{(via bounding box 
comparison)}.
If so, the list of \emph{periodic faces} of the
bounding box $\boundingBox$ of $\domain$ along which $\domain_i'$ is located is
identified (i.e. left, right, bottom, top, front, back).
\major{This information is used to trigger exchanges of data chunks, as 
illustrated in \autoref{fig_periodicCondition}\emph{(c)}, whose extent depends 
\discuss{on} the periodic face type (corner, edge, face).}
Additionally, the local adjacency graph 
\major{is updated to account for blocks which are adjacent via the periodic 
boundaries}.

\major{Similarly to \autoref{implicitTriangulation}, runtime queries are 
performed on each process by querying the local copy of the global periodic 
triangulation $\domain$ (with the necessary local-to-global identifier 
translations).}


%% file: distributing_ttk.tex
\section{Distributed Pipeline}
\label{sec_ttkDistributedInfrastructure}

This section provides an overview of the overall processing by TTK of a
distributed dataset.
\major{It} documents
the
preconditioning steps
handled by the core
infrastructure of TTK
\major{(beyond the triangulation handling, \autoref{sec_distributedTriangulation})}
in order to complete the
\major{support}
of the distributed model
\major{specified}
in
\autoref{sec_distributedSetup}.
\major{We refer the reader to Appendix A.1 for further details
on
the
integration of TTK with VTK and ParaView.}

\begin{figure}
  \includegraphics[width=\linewidth]{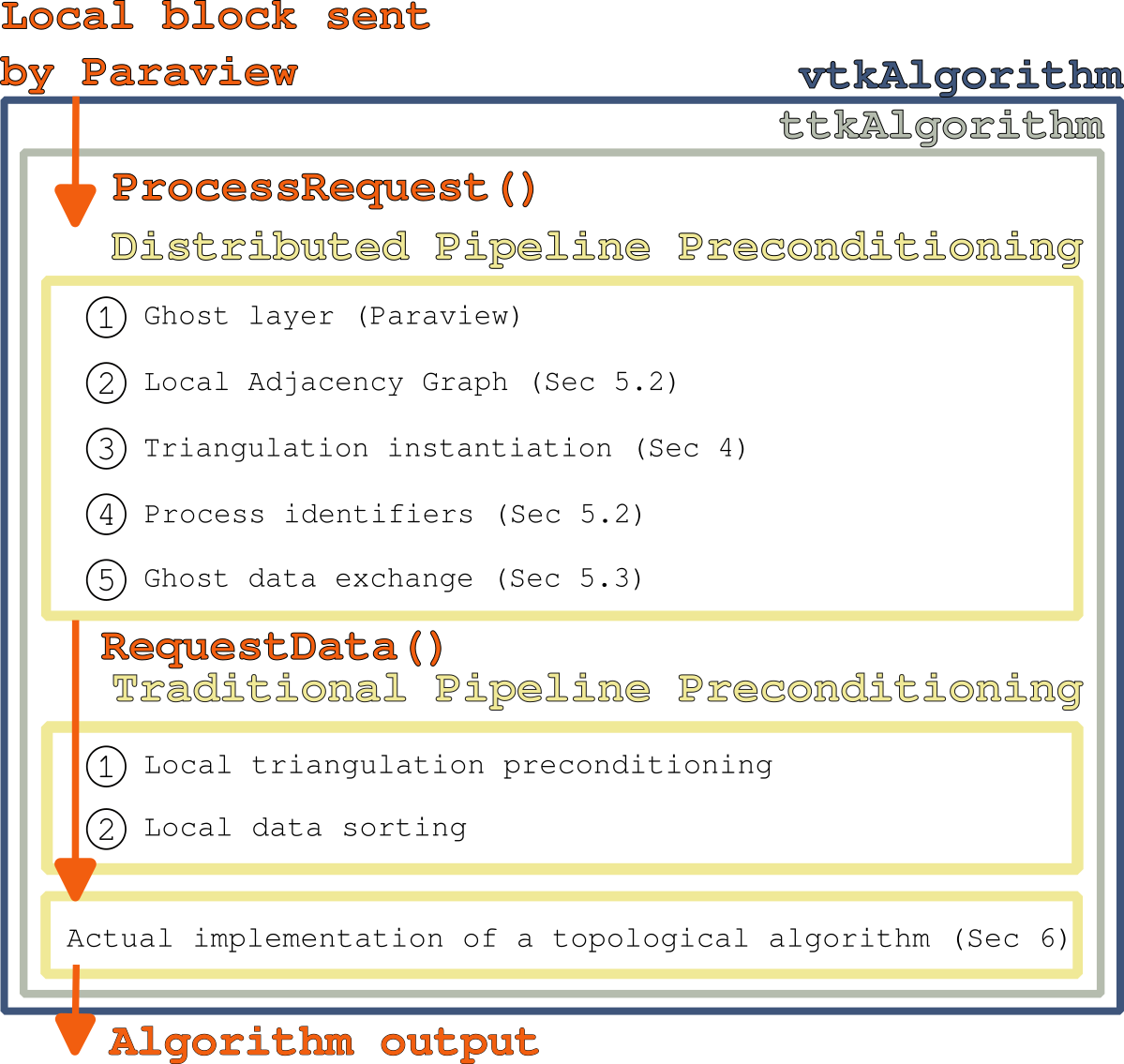}
  \caption{Overview of the overall pipeline
  upon the the
delivery of a data block $\domain_i$ by ParaView (top).
A step of pipeline preconditioning specialized for the distributed
setting (top \eve{yellow} frame) is automatically triggered before calling the actual
implementation of the topological algorithm.
Note that
each
preconditioning phase
is only triggered if the
corresponding information has not been cached yet. Then, for practical
pipelines, the preconditioning typically only occurs before the first
algorithm of the pipeline.
}
  \label{fig_schematicOverview}
\end{figure}

\subsection{Overview}
\label{ref_pipelineOverview}
The input data is
provided in the form of a distributed dataset
\major{(see \autoref{sec_distributedInputFormalization})}
loaded from \eve{a
filesystem} (e.g. \emph{PVTI} file format) or provided in-situ \eve{(e.g. with
\emph{Catalyst})}.
\major{As shown in \autoref{fig_schematicOverview} (and detailed
in Appendix A.1), ParaView's execution flow enters TTK via the function}
\verb|ProcessRequest|\major{, which triggers TTK's
preconditioning,
including the \emph{Distributed Pipeline Preconditioning} (specific to
the distributed mode, top yellow frame) prior to
the traditional, local preconditioning (middle yellow frame) and
finally the implementation of the topological algorithm  (bottom yellow
frame).
In the following, we describe the \emph{Distributed Pipeline Preconditioning}.}


\noindent
\textbf{\emph{(1)} Ghost layer generation:} if the local data block does not
include any ghost cells, the ghost layer generation algorithm (implemented by
ParaView) is automatically triggered. This step is omitted if a valid ghost
layer is already present.

\noindent
\textbf{\emph{(2)} Local adjacency graph (LAG) initialization:}
\major{An}
estimation of the \major{local} adjacency graph
(i.e.
connecting \major{the data block} to
its \major{neighbors}) is constructed.
\major{This step (described in \autoref{sec_highLevelFrameWork}) is omitted if a
valid LAG is
already present.}

\noindent
\textbf{\emph{(3)} Triangulation instantiation:} this step instantiates a new
TTK 
triangulation data structure (\autoref{sec_distributedTriangulation}).
This step is omitted if a valid triangulation is already present.

\noindent
\textbf{\emph{(4)} \majorRevision{Simplex-to-process map}
generation:} this step computes the
\majorRevision{simplex-to-process} identifier for each simplex (as
specified in
\autoref{sec_distributedInputFormalization}).
This step
\major{(described in \autoref{sec_highLevelFrameWork})} is omitted if valid
\majorRevision{simplex-to-process maps} are already present.

\noindent
\eve{\textbf{\emph{(5)} Ghost data exchange:} this step computes for each neighbor process
\major{$j$}
the list of vertices or cells
\major{exclusively}
owned by
\major{it, and}
which are ghosts
in the
process \major{$i$}.
This step \major{(described in \autoref{sec_lowLevelInterface})} is optional and
is only triggered
if the calling algorithm pre-declared its usage
\major{at}
preconditioning.}

After these steps, the traditional TTK preconditioning is executed
\major{(middle yellow frame, \autoref{fig_schematicOverview})}.

\subsection{\major{Infrastructure details}}
\label{sec_lowLevelInterface}
\label{sec_highLevelFrameWork}
\label{sec_highLevelInterface}
\label{sec_rankArrays}
This section describes the implementation of the pipeline
preconditioning mentioned in the above overview
(\autoref{ref_pipelineOverview}), specifically, the routines which are not
directly related to the distributed triangulation (which has been covered in
\autoref{sec_distributedTriangulation}).

\noindent
\textbf{Local adjacency graph (LAG) initialization:} Given a ghosted
block $\domain_i'$, the goal of this step is to store a list of processes,
which are responsible for the blocks adjacent to $\domain_i'$
(\autoref{fig_distributedModel}\emph{(e)}).
First, each process $i$ computes the bounding box $\boundingBox_i$ of
its ghosted block $\domain_i'$. Next, all processes exchange their bounding
boxes.
\major{Finally,}
each process \major{$i$} can initialize a list of neighbor processes by
collecting the processes whose bounding box intersects with
$\boundingBox_i$.
This first estimation of the LAG will be \major{refined
(next paragraph)}
\eve{after} the generation
of the \majorRevision{simplex-to-process} identifiers
\major{(which is relevant in
the case of explicitly triangulated domains).}

\noindent
\textbf{\majorRevision{Simplex-to-process map} generation:}
As
\major{specified}
in \autoref{sec_distributedInputFormalization},
\majorRevision{each simplex
is associated to
the identifier of the process which exclusively owns it.}
This
convenience
feature can be particularly useful to quickly identify where to continue a
local processing when reaching the boundary of a block (e.g. integral lines,
\major{\autoref{sec_integralLines}}).

\noindent
Each vertex $v \in \domain_i'$ is
\major{classified}
by ParaView as
ghost
\major{or non-ghost.}
For each
\major{non-ghost}
vertex $v$,
we set
\major{its simplex-to-process identifier to $i$.}
Then,
\major{the \emph{ghost global identifier list} is computed (it contains the global identifiers of all the ghost vertices of $\domain_i'$).
%
%
Next, this list}
is sent to each process $j$
marked as being adjacent in the LAG (previous paragraph). Then, the process $j$
will return its
identifier ($j$)
\major{and the subset of the \emph{ghost global identifier list}, corresponding to non-ghost vertices in $\domain_j'$.}
%
Finally, the process
$i$ will set the
\major{simplex-to-process identifier of $v$ to $j$,}
for each
\major{vertex $v$ returned}
\eve{by
$j$. The} procedure for the $d$-simplices is identical.
\majorRevision{The simplex-to-process maps} for the
simplices of intermediate dimensions
\major{are}
inferred from these of the $d$-simplices,
\major{as specified in}
\autoref{sec_distributedInputFormalization}.
\major{Following}
the generation of \majorRevision{the simplex-to-process maps,}
the LAG is
\major{updated,
by only considering the block $i$ and $j$ as neighbors if $i$
contains ghost vertices which are exclusively owned by $j$ and reciprocally.}

\eve{In implicit
\major{mode,}
the preconditioning
of
\majorRevision{the simplex-to-process map}
generation is limited to
the computation of
discrete bounding boxes \major{(i.e. expressed in terms of global discrete coordinates)}
\major{for}
the non-ghosted block $\domain_i$.
The bounding boxes
are then exchanged between neighboring processes. \major{Then, the}
\majorRevision{simplex-to-process maps}
\major{are}
inferred
on-the-fly, at
query-time, from the discrete bounding boxes.}


\noindent
\textbf{Ghost data exchange:}
In many scenarios, it may be desirable
to update the data attached to the ghost simplices of a given block
$\domain_i'$.
\major{For instance, when considering smoothing (\autoref{sec_smoother}), at}
each iteration, the process $i$ needs to
retrieve the new, smoothed $f$ data values for its ghost vertices, prior to
\major{the next smoothing iteration.}
We implement this task in TTK as a simple convenience
function. \eve{\major{First, using}
the list of neighbors (collected from the LAG),
\julien{the}
process $i$ will,
for each neighbor process $j$, send the global identifiers of the simplices
\julien{which are}
ghost for
$i$ and owned by $j$ (using \major{their \majorRevision{simplex-to-process maps).}}
\major{This is computed once,} in an optional preconditioning step (step 5, \autoref{ref_pipelineOverview}).
\major{This}
list of ghost vertex identifiers
\julien{is}
cached \major{in $j$} and used at runtime\major{, when necessary,}
to send \major{to $i$} the
updated values \major{(exchange data buffers are updated with shared-memory parallelism)}.}
A similar procedure is available
for $d$-simplices.

%% file: examples.tex
\section{Examples}
\label{sec_examples}

Secs. \ref{sec_distributedTriangulation} and
\ref{sec_ttkDistributedInfrastructure} documented the implementation of the
distributed model specified in \autoref{sec_distributedSetup}. In this section,
we now describe how to make use of this model to
\julien{extend}
topological algorithms to
the distributed setting. Specifically, we will mostly focus on the algorithms
described in \autoref{sec_background}.

\subsection{Algorithm taxonomy}
\label{sec_taxonomy}
In this section, we
\julien{present}
a taxonomy of the topological algorithms
implemented in TTK, based on their
needs of communications
\pierre{on distributed-memory architectures}.

\noindent
\textbf{\emph{(1)} No Communication (NC):} This category includes algorithms
for which processes do not need to communicate with each other to complete
their computation. This is the simplest form of algorithm\pierre{s} and
the easiest to
\julien{extend}
to a distributed setting. Such algorithms are often referred
to as \emph{\pierre{embarrassingly} parallel}. In TTK, this includes
algorithms performing local operations and generating a local output,
e.g.:
critical
point classification \autoref{sec_criticalPoints}, discrete gradient computation
\autoref{sec_discreteGradient}, Jacobi set extraction \cite{jacobiSets}, Fiber
surface computation  \cite{KlacanskyTCG17} \eve{and}
marching tetrahedra.

\noindent
\textbf{\emph{(2)} Data-Independent Communications (DIC):} This category
includes algorithms for which processes do need to communicate with each other,
but at predictable stages of the algorithm, \pierre{with a predictable set of
processes \eve{and communication} volume}, independently
of \eve{the data} values. This typically corresponds
to algorithms \eve{performing} a local operation on their block \eve{that} need
intermediate results from adjacent blocks to finalize their computation. In TTK,
this includes for instance: data normalization, data or geometry smoothing
(\autoref{sec_exampleImplementations}),
or continuous scatter plots
\cite{BachthalerW08a}.


\noindent
\textbf{\emph{(3)} Data-Dependent Communications (DDC):} This category includes
algorithms which do not fall within the previous categories, i.e. for which
communications can occur at unpredictable stages of the algorithm, \pierre{with an
unpredictable set of \eve{processes or communication}
volume}, depending on the data values. This
is the most difficult category of algorithms to
\julien{extend}
to the distributed
setting, 
\julien{since}
\pierre{an efficient port would require} a complete re-design of the
algorithm. Unfortunately, we conjecture that
most
topological algorithms fall into that category.
In TTK, this includes for instance:
integral lines (\autoref{sec_integralLines_def}),
persistence
diagrams \cite{guillou_tech22}, merge and contour trees \cite{gueunet_tpds19}, path compression \cite{robin23},
Reeb graphs \cite{gueunet_egpgv19}, Morse-Smale complexes \cite{ttk17},
Rips complexes, topological simplification \cite{tierny_vis12,
Lukasczyk_vis20}, Reeb spaces \cite{tierny_vis16}, etc.

%
%

\begin{figure*}
  \centering
  \includegraphics[width=\linewidth]{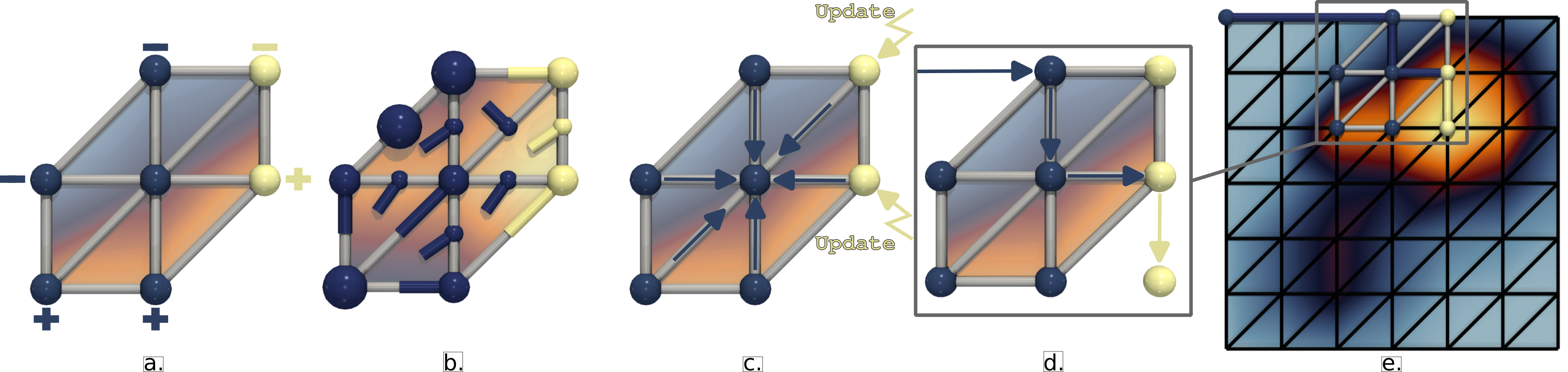}
  \caption{\eve{Examples of topological algorithm modifications for the support of distributed memory computation.
  \emph{(a)} Scalar Field Critical Points (NC): Critical points are generated similarly to the sequential mode. Upper and lower \julien{links} ($+$ and $-$ signs in the figure)  of non-ghost vertices on the boundary of $\domain_i$ are computed using ghost vertices (here in yellow).
  \emph{(b)} Discrete Gradient (NC): Similarly to (a), this algorithm processes each vertex of the domain independently. For each non-ghost vertex on the boundary of $\domain_i$, the lower link
  computation can rely on ghost vertices. Critical simplices are represented by bigger spheres.
  \emph{(c)} Scalar Field Smoother (DIC): This procedure smooths a scalar field $f$ by local
averaging for a user-defined number of iterations. The values of ghost vertices (in yellow) will need to be updated after each iteration.
  \emph{(d)} and \emph{(e)} Integral Lines (DDC): \emph{(e)} each process will compute the integral lines whose seeds lie within its block $\domain_i$.
  Then either the integral line reaches its final vertex within $\domain_i$, \emph{completing} the computation, or the integral line
reaches a vertex outside of $\domain_i$ (here in yellow in \emph{(d)}). In the latter case,
the integral line data is stored to be sent later to the yellow process.
Once all the work is done on all processes, they exchange the data of incomplete integral lines
and resume \julien{the} computation of the integral lines on their block. The computation stops when all integral lines have completed.}}
  \label{fig_distributedExamples}
\end{figure*}

\input{mpi+threads.tex}

\subsection{\pierrre{Distributed algorithm examples}}
\label{sec_exampleImplementations}
We now illustrate the
taxonomy
\major{of}
\autoref{sec_taxonomy}
by describing the
\pierrre{distributed-memory parallelization} 
of algorithms
belonging to each of the categories, \pierrre{while} 
exploiting the distributed model
we introduced (\autoref{sec_distributedSetup}).


\subsubsection{NC: Scalar Field Critical Points}
\label{sec_scalarFieldCriticalPoints}
This algorithm processes each vertex $v$ of the domain independently
and
\major{performs}
the classification presented in
\autoref{sec_criticalPoints}.
Since it processes a local piece of data (the lower and upper links
$\Link^-(v)$ and $\Link^+(v)$) and that it generates a localized output (a list
of critical points for the local block), it does not require any communication \eve{(\autoref{fig_distributedExamples}\emph{(a)})}.
Thus, it is classified in the category NC of the above taxonomy.
To port this \pierre{embarrassingly} parallel algorithm to the
distributed setting, two modifications are required.
First, the algorithm \eve{does not classify} ghost vertices (which will be classified by other processes).
Second, to fulfill the distributed output specification
(\autoref{sec_distributedOutput}), each output critical point is associated
with its \emph{global} vertex identifier (instead of its local
\major{one).}


\subsubsection{NC: Discrete Gradient}
Similarly to the previous case, this algorithm processes each vertex $v$ of the
domain independently. Specifically, it generates discrete vectors for the
lower star
$\Star^-(v)$
and the simplices which are assigned to no discrete vectors
are stored as critical simplices (\autoref{sec_discreteGradient}).
Similarly to the previous case, this algorithm only requires local data and
only produces local outputs, without needing communications (hence its NC
classification) \eve{(\autoref{fig_distributedExamples} \emph{(b)})}.
The port of
this \pierre{embarrassingly} parallel algorithm requires \eve{two modifications.
First, only the vertices which are exclusively owned by
the current process (\autoref{sec_distributedInputFormalization}) are processed. 
The gradient \julien{for} ghost vertices, and the simplices in their lower links, is not computed.
Second,}
similarly to the previous case, the simplex identifiers
associated with the discrete vectors and critical simplices are expressed with
\emph{global} identifiers
(instead of local ones).

%

\subsubsection{DIC: Scalar Field Normalizer}
This
procedure
normalizes an input scalar field $f$ to the
range $[0, 1]$.
\major{It is}
divided into two steps. \eve{First, each process
\pierre{computes} 
its local extreme values and all processes exchange their extreme
values to determine the values $f_{min}$ and $f_{max}$ for the entire domain $\domain$ 
\pierre{using MPI collective communications.}}
Second, all data values are normalized independently, based on $f_{min}$ and
$f_{max}$. The first step of the algorithm requires inter-process communications
in a way which is predictable and independent of the actual data values (hence
its DIC classification in the taxonomy). 


 \subsubsection{DIC: Scalar Field Smoother}
 \label{sec_smoother}
 This
 procedure
 smooths a scalar field $f$ by local
averaging (i.e. by replacing $f(v)$ with the average data values on the
vertices of $\Star(v)$). This averaging procedure is typically iterated for a
user-defined number of iterations. However, at a given
\major{iteration,}
in order
to guarantee a correct result for each vertex $v$
located on the boundary of the
local block (i.e. $v$ is a non-ghost vertex adjacent to ghost-vertices), the
updated $f$ values from the previous iteration
need to be retrieved for
each of its ghost neighbors \eve{(\autoref{fig_distributedExamples}\emph{(c)})}. Thus, at the end of each iteration, each
process $i$ needs to communicate with its neighbors to retrieve the smoothed
values for its ghost vertices, which is achieved by using the generic ghost
data exchange procedure described in \autoref{sec_lowLevelInterface} (hence
the DIC classification for this algorithm).

%

 \subsubsection{DDC: Integral lines}
\label{sec_integralLines}
\eve{
Unlike the previous cases,  the port of this algorithm requires quite extensive modifications. The first step is similar to its sequential version (Sec \ref{sec_integralLines_def}):
each process
\major{$i$}
will compute the integral lines whose seeds lie within its block
$\domain_i$ \major{(each seed is processed independently via shared-memory 
parallelism with OpenMP)}. 
\majorRevision{Moreover, the process $i$
will be marked as the exclusive owner
of the part of the integral line (i.e. the
vertices and edges of the sub-geometry) created on its block.} From there, two possibilities arise: \pierre{either} the integral line reaches its final vertex
within $\domain_i$, \emph{completing} the computation, or the integral line
 reaches a ghost vertex owned by another process $j$ and is \emph{incomplete}. In the latter case,
some of the integral line data (such as global identifier, the distance from the seed or the global identifier of the seed) is stored in a vector to be sent
\major{later}
to \major{the} process $j$
\eve{(\autoref{fig_distributedExamples}\emph{(d) and \emph{(e)}})}.
Once all integral lines on all processes are marked as either complete or incomplete,
all processes exchange the data of their incomplete integral lines
and use that data to resume
computation of the integral lines on their block.}

\eve{
These computation and communication steps are run until all integral lines on all processes
are completed. Consequently, depending on the dataset, \pierre{and
  the process}, there may be very little
communication, \pierre{e.g.} if all the integral lines \pierre{lie} 
within the bounds of \pierre{a block},
or a lot of communications, \pierre{e.g.} if \pierre{some} 
integral lines are defined across the
blocks of multiple processes (hence its DDC classification).
}


 \begin{figure}
    \centering
    \includegraphics[width=\linewidth]{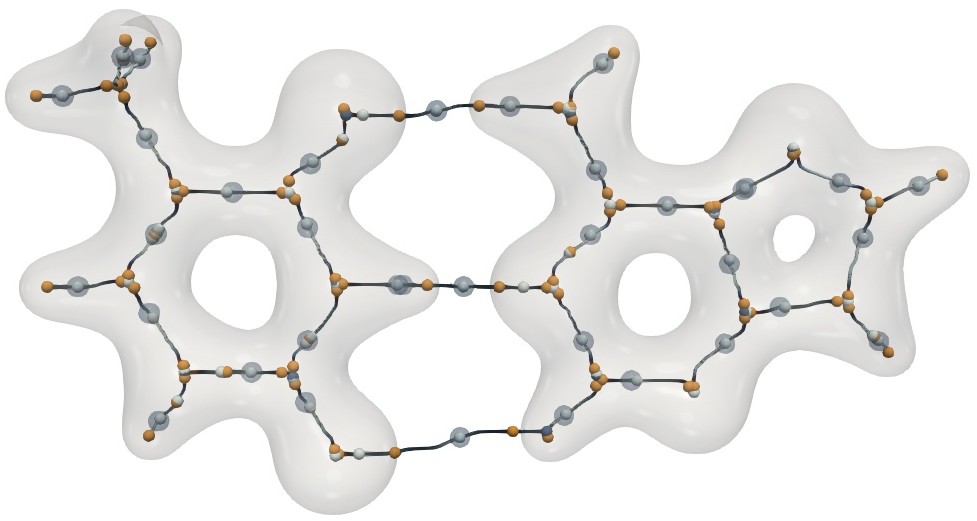}
    \caption{\eve{Output of the integrated pipeline on the AT dataset, a three-dimensional regular grid
\julien{of}
            the electronic density \julien{(and its gradient magnitude)} in the Adenine Thymine complex (AT). The extracted integral lines capture the covalent and hydrogen bonds within the molecule complex.
            The transparent spheres are the critical points used as seeds of the integral lines while
            the full spheres are the critical points
            \julien{of}
            $\julien{|\nabla f|}$ and
            show where the electronic density experiences rapid changes, indicating transition points occurring within the bond.
            This image was obtained by resampling the original dataset to 
            \majorRevision{$2048^3$}
            and executing the integrated pipeline on 64 nodes of 24 
cores each 
(\majorRevision{1536}
cores) on MeSU-beta.}}
    \label{fig_atExample}
  \end{figure}

\subsection{Integrated pipeline}
\label{sec_integratedPipeline}

\begin{table}
\begin{center}
\begin{tabular}{|lll|c@{}}
 \hline
 Abbreviation  & Algorithm & Input \\
 \hline
 \julien{1.} SFS1 & ScalarFieldSmoother & $f$\\
 \hline
 \julien{2.} SFS2 & ScalarFieldSmoother & $|\julien{\nabla} f|$ \\
 \hline
 \julien{3.} SFN1 & ScalarFieldNormalizer &  $f_{SFS1}$\\
 \hline
 \julien{4.} AP & ArrayPreconditioning & $f_{SFN1}$\\
 \hline
 \julien{5.} SFCP1 & ScalarFieldCriticalPoints & $f_{AP}$ \\
 \hline
 \julien{6.} IL & IntegralLines & \makecell[l]{$f_{AP}$ (domain), \\ $f_{SFCP1}$ (seeds)} \\
 \hline
 \julien{7.} GS & GeometrySmoother & $f_{IL}$\\
 \hline
 \julien{8.} SFCP2 & ScalarFieldCriticalPoints & $|\julien{\nabla} f|_{SFS2}$ on $\domain_{GS}$\\
 \hline
\end{tabular}
\end{center}
\caption{\eve{Composition of the integrated pipeline. 
Each line denotes an algorithm in the pipeline, by order of appearance (top to bottom), as well as its input.
$f$ is the input scalar field. Each algorithm modifies the scalar field: $f_{A}$ is 
the modified scalar field $f$\julien{,} output of algorithm $A$. $\domain_{GS}$ is the output
domain of GeometrySmoother.}}
\label{table_integratedPipeline}
\end{table}

\eve{In this section, we describe an integrated pipeline that produces a real-life use case
\julien{combining all the}
the port examples presented in
Sec.\ref{sec_exampleImplementations}. All of the algorithms, their order as well as their input are described in Table \ref{table_integratedPipeline}.
The input dataset is a three-dimensional regular grid with two scalar fields $f$,
the electronic density in the \emph{Adenine Thymine} complex (AT) and its \julien{gradient} magnitude $|\julien{\nabla} f|$.
First, $f$ and $|\julien{\nabla}f|$ are smoothed and $f$ is normalized. Critical points
\julien{of} $f$
are computed and used as seeds
\julien{to compute}
integral lines \julien{of} $f$. The
extracted integral lines capture the covalent and hydrogen bonds within the 
molecule complex \julien{(\autoref{fig_atExample})}. 
\majorRevision{Then,} critical points are computed for $|\julien{\nabla}f|$ on
the
integral lines. The extracted critical points indicate locations of covalent bonds 
where the electronic density experiences rapid changes, indicating transition points occurring 
within the bond \julien{(\autoref{fig_atExample})}.}

\eve{The local order of $f$ is required by two algorithms: the first critical points (SFCP1) and the integral lines (IL).
\julien{Since these two algorithms are separate leaves of the pipeline, each of them would trigger the automatic local order computation. Instead, to avoid this duplicated computation, we manually call the local order computation in a preprocess (i.e. by calling the \emph{ArrayPreconditioning} algorithm).}}
%

The chosen
dataset is intentionally quite small ($177\times95\times48$) to ensure
reproducibility. It is resampled before the pipeline
to create a more sizable
example, using \major{ParaView's} \julien{\emph{ResampleToImage}} \major{feature}
\major{(i.e. grid resampling via trilinear interpolation).}
Anyone can execute this pipeline to the best of their resources, by choosing the appropriate
resampling dimensions. In our case, the new dataset is of dimensions 
\majorRevision{($2048^3$),}
encompassing roughly 8.5 billion vertices.

\eve{The pipeline was also run on a second, larger, dataset \julien{(\emph{Turbulent Channel Flow})}, to show TTK's capability to handle massive datasets \julien{(specifically, the largest publicly available dataset we have found)}.
\julien{This dataset represents}
a three dimensional pressure field of a direct numerical simulation of a fully developed flow at different
Reynolds numbers in a plane channel (obtained from the Open Scientific Visualization Datasets \cite{openSciVisDataSets}).
\julien{Its dimensions are}
\majorRevision{($10240\times7680\times1536$),}
which
\julien{is}
approximately 120 billion vertices.
Before applying the pipeline,
\julien{the gradient magnitude is computed and added to the dataset, and the result is}
converted
\pierre{using single-precision floating-point numbers}
\julien{(thereby reducing memory consumption \pierre{at runtime}).
}}

%% file: mpi+threads.tex

\subsection{Hybrid MPI+thread strategy}
\label{s:MPI+thread}
In this section, we present general aspects regarding the combination
of \discuss{multi-process parallelism (with MPI; running on both 
  distributed- and shared-memory architectures) and multi-thread
  parallelism (restricted to shared-memory architectures)}, that
are
used within the
examples
(\autoref{sec_exampleImplementations}).

Current compute clusters are based on multiple nodes, each node
\julien{including}
multi-core processors.
For performance reasons, one then \majorRevision{runs} one execution flow per
core
following either a {\em pure MPI} strategy, i.e. with one MPI process per
core, or a {\em MPI+thread} strategy, i.e. with one MPI process per
node (or per processor) and multiple threads within each MPI process.
The latter can
\julien{improve performance}
thanks to fewer MPI
communications (due to fewer MPI processes), to a (better) dynamic
load balancing among threads within each MPI process, and to
a multi-core speedup for computations specifically performed by the MPI process 0
(e.g. Sec. 4.2.1).
The overall memory footprint is also lower with the latter, since
using fewer MPI processes implies fewer ghost simplices and
less data duplication.

Regarding the MPI+thread strategy and the port examples
  described in
  \autoref{sec_exampleImplementations},
  we rely in TTK on the
\verb|MPI_THREAD_FUNNELED|
thread support level in MPI \cite{mpi40}.
  According to this level of thread support, only the master
  (i.e. original) thread can issue calls to MPI routines.
In each port example, within each MPI process, the
communication steps (if any) are thus
performed in serial whereas the computation steps are multi-threaded, using the OpenMP implementations
already available in TTK \major{(Appendix A.3)}.

Besides, when using the MPI+thread implementations, one can
  choose to run one MPI process per node or one MPI process per processor,
  hence e.g. two MPI processes on a node with two processors.
  The former leads to fewer MPI processes in total, and enables one to
  balance the compute load among all the cores of the node. The latter
  can avoid performance issues due to NUMA (non-uniform memory access)
  effects which occur when a thread running on a given processor accesses data on
  the memory local to the other processor.
  Options specific to the MPI implementation enable the user to
  choose one of the two possibilities.
  The threads are also bound to the CPU cores using the
  OpenMP thread affinity features \cite{openmp51}.

%% file: results.tex
\section{Results}
\label{sec_results}

For the following results, we rely on Sorbonne Université's
supercomputer, MeSU-beta. MeSU-beta is a 
\pierre{compute cluster} with 144 nodes of 24 cores each
(\pierre{totaling 
\majorRevision{$3456$}
cores}). 
\eve{Its nodes are composed of 2 Intel Xeon E5-2670v3 (2.7 GHz, 12
  cores), \pierre{with SMT (simultaneous multithreading) disabled
    (i.e. running 1 thread per core), and} with 
128GB of memory each. The nodes are interconnect with Mellanox Infiniband.}

When measuring the performance of
\major{a specific}
algorithm, only the execution of the algorithm \majorRevision{itself}
is timed \major{(using the timing method described in Appendix B)}. None of the
preconditioning or input and output formatting is timed unless 
\eve{explicitly} stated. \eve{The preconditioning steps are an investment in time: they can be used again by 
other algorithms later on in the pipeline, thus, including the cost of these steps in the execution time 
of a single algorithm would not provide an accurate representation of performance in a more
complicated pipeline. They are therefore excluded from the
\julien{individual}
benchmarks \julien{(\autoref{sec_distributedAlgoPerf})}
\julien{but}
included
in the study of the \julien{global,} integrated pipeline 
\julien{\autoref{sec_pipeline}}
\major{(which is timed using ParaView's internal timer)}.}


The benchmark is performed on five
different
datasets: \eve{\emph{Wavelet} (3D wavelets on a cube),
\emph{Elevation} (synthetic dataset of the altitude within a cube, with a unique maximum at
one corner of the cube and a unique minimum at the opposite corner), \emph{Isabel}
(magnitude of the wind velocity in a simulation of the hurricane Isabel that hit the
east coast of the USA in 2003), \emph{Random} (random field on a cube) and \emph{Backpack}
(density in the CT-scan of a backpack filled with items).}
The datasets
\julien{all}
originate  from
publicly available
\julien{repositories} \cite{openSciVisDataSets, ttkData}.

\subsection{\eve{Distributed algorithms performance}}
\label{sec_distributedAlgoPerf}

\major{This section evaluates the practical performance of
the extension to
the distributed setting (\autoref{sec_exampleImplementations}) of
the
algorithms presented in  \autoref{sec_background}, by
\discuss{considering strong and weak scaling.}
%
}


\subsubsection{\majorRevision{Strong scaling}}
\label{s:res_strong}
\begin{figure}
    \centering
    \includegraphics[width=0.5\textwidth]{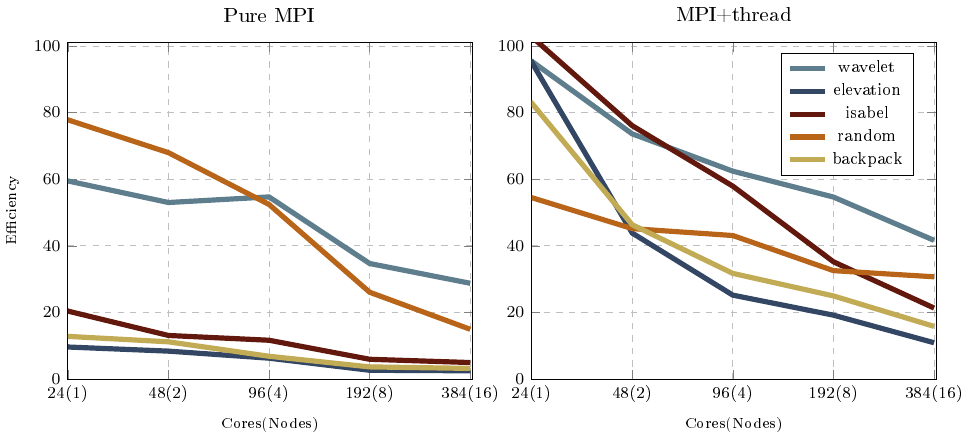}
    \caption{\pierre{\major{Strong scaling 
    efficiencies} for the Integral line \julien{computation}
        algorithm \eve{with $500,000$ seeds, randomly distributed on all processes,} using the pure MPI strategy (left) and the MPI+thread one (right) with
        1 MPI process and 24 threads per node.}  
      \pierre{The MPI+thread strategy is significantly more efficient than
        the pure MPI one.}}
    \label{fig_benchmarkIntegralLines}
  \end{figure}

\begin{figure*}
    \centering
    \includegraphics[width=\linewidth]{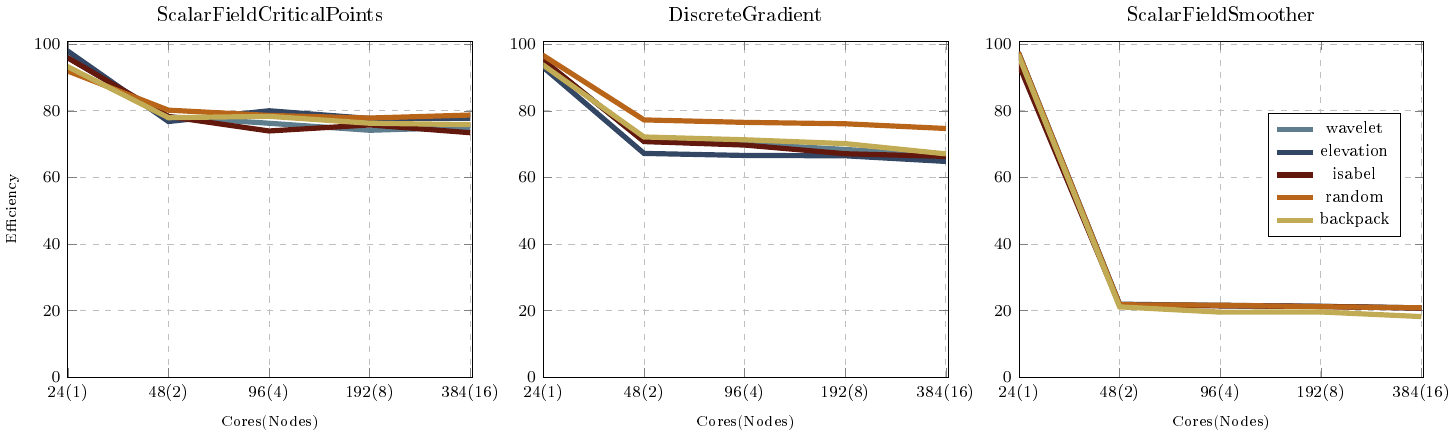}
    \caption{\pierre{
    \major{Strong scaling efficiencies}
    for various
algorithms
        \major{(MPI+thread: 1 MPI process and 24 threads per node).}}}
    \label{fig_benchmarkOther}
  \end{figure*}

\majorRevision{For a given problem size, we first evaluate the runtime performance of our novel framework for distributed computations in TTK, as more computational resources are available. For this,
we conduct a strong scaling analysis, where the size of
the input data is constant
(each dataset is resampled to $512^3$ via trilinear interpolation)
while the number of available cores increases.
The speedup $s_p$ for $p$ cores is defined as $ s_p = \frac{t_1}{t_p}$, with $t_p$ being the execution time 
for $p$ cores. 
Then, we define the \emph{strong scaling efficiency}
for $p$ cores 
as $\frac{s_p}{p} \times 100$. Appendix
\discuss{C} 
shows the same results as presented in Fig. \ref{fig_benchmarkIntegralLines} (right) and Fig. \ref{fig_benchmarkOther}, but in terms of execution time instead of
parallel
efficiency.}

\pierre{We first compare the pure MPI and the MPI+thread strategies
(\autoref{s:MPI+thread}).
Regarding the MPI+thread strategy, we rely
on one MPI process per node (and 24 threads within) instead of one MPI
process per processor (and 12 threads each).
According to
 performance tests (not shown here), 
both options lead indeed to
similar performance results, except when using one single node: in
this case, having one single MPI process (no communication and no
ghost simplices required) is more efficient than two.
Having
one MPI process per node \pierre{also leads to} a
lower memory
\major{usage.}
}

\pierre{As shown in Fig. \ref{fig_benchmarkIntegralLines}, using
  MPI+thread  (with one MPI process and 24 threads per node) 
  is then substantially more efficient than using a pure MPI design for
  the integral line algorithm, for all datasets except the \emph{Random} dataset. 
  More precisely, even for MPI+thread, the efficiency decreases 
  with the number of cores and depends significantly on the dataset.
  This is due to a strong 
  workload imbalance between the processes: the integral lines are not
  evenly distributed on the MPI processes which can lead to long idle
  periods for some processes (waiting for the other to process their
  integral lines). This applies to the \emph{Backpack} dataset for example.
  Regarding the \emph{Elevation} dataset (very smooth,
with only one maximum and one minimum) or the \emph{Isabel} one (very smooth
too), the generated integral lines are here especially lengthy and span
several (but not all) processes, leading to low efficiencies. 
On the contrary the \emph{Random} dataset is very balanced, but is also very noisy, leading to
very short integral lines: for the same number of integral lines, the
computation times are much shorter than for the other datasets which
makes the communication cost more detrimental to performance.
Finally, the \emph{Wavelet} dataset is the most balanced one, with long enough
integral lines, and thus shows the best performance results.
Compared to the pure MPI strategy, the \eve{MPI+thread one} benefits 
from fewer MPI processes and therefore from a lower load imbalance.}

\pierre{\major{We have tried}
  to improve the parallel efficiencies of the integral line
  algorithm, by dedicating a thread to MPI communications.
  Thanks to this thread,} \eve{ an incomplete integral line is sent right away, without waiting for all integral lines on the 
process to be computed. Each process also continuously receives integral lines and adds them
immediately to the pool of integral lines to be computed.} 
\pierre{However this design based on a communication thread adds a significant
  amount of complexity to the implementation (due to the required thread synchronizations),
  and did not \majorRevision{improve the parallel efficicency} since the main
  performance bottleneck is the load imbalance among processes.
  As a result, we do not rely on this communication thread design in
  our distributed integral lines implementation.}

\eve{The performance \pierre{results} for the other distributed algorithms can be found 
in \autoref{fig_benchmarkOther}. For the \emph{ScalarFieldCriticalPoints},
\pierre{a very} good efficiency \julien{($80 \%$)} is achieved \julien{(which is comparable to
its shared-memory parallel implementation on one node, \julien{$90 \%$})}, with little dependence on the dataset.
The \emph{DiscreteGradient} likewise performs very well in terms of efficiency,
albeit slightly less, 
due to the parallelization method of the algorithm, for which
adding ghost simplices will add a small amount of \pierre{extra work
  in parallel.
These two algorithms strongly benefit from parallel computing, even when using hundreds of
cores.}
The \emph{ScalarFieldSmoother} exhibits
lower efficiency. 
This can be explained by the \pierre{need for} communications at each iteration, 
as well as by the \pierre{low cost of the smoothing process
(which is a simple averaging operation).}
Indeed, the faster a \pierre{computation}, the
\pierre{stronger the impact of communications on the overall
  performance}.
}

\discuss{
Finally we emphasize that, at the exception of \emph{IntegralLines} (for which we derived a new
implementation, \autoref{sec_integralLines}), shared-memory parallel
implementations of these algorithms (using OpenMP
threads) pre-existed in TTK prior to this work.
In our MPI+thread strategy, 
we leverage these same shared-memory parallel implementations
regarding multi-thread parallelism. 
Moreover when using only one MPI process, MPI communications are not
triggered and processing specific to the distributed setting (e.g. on
ghost simplices)  is not carried out. Thus, when running our
novel MPI+thread extension of these algorithms on only one MPI
process,
performances are identical to these of the pre-existing,
shared-memory-only implementations.
}

\subsubsection{\majorRevision{Weak scaling}}
\begin{figure}
  \centering
  \includegraphics[width=\linewidth]{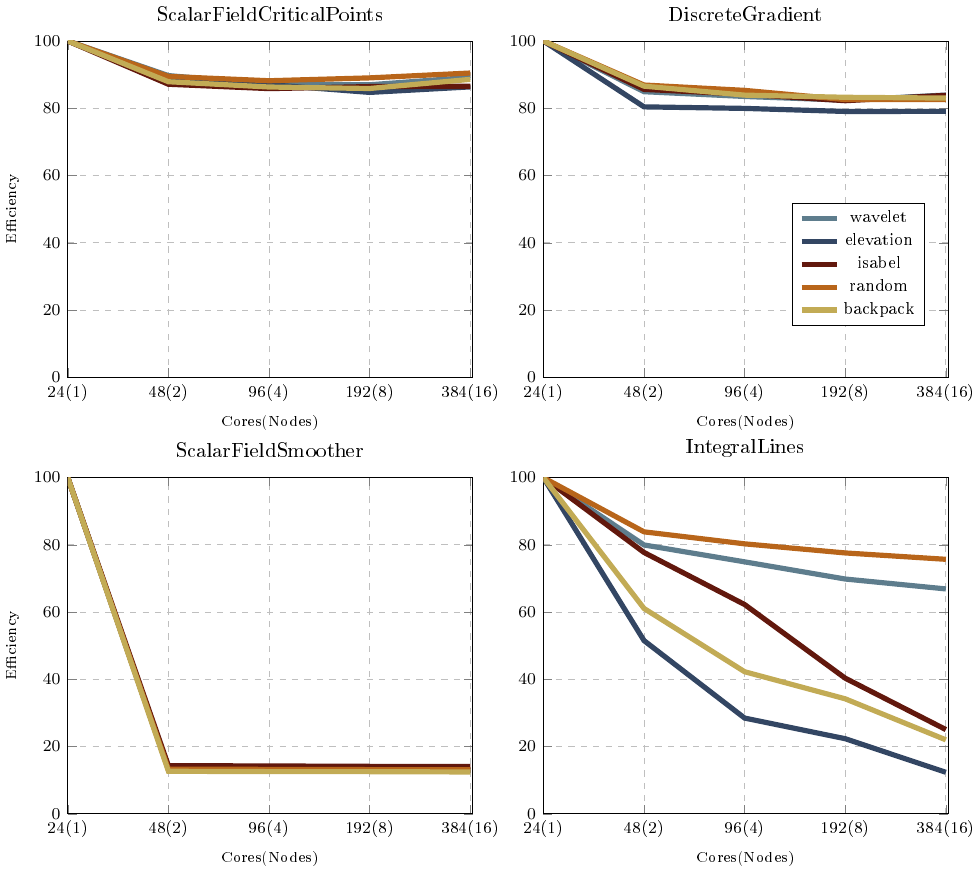}
  \caption{\majorRevision{Weak scaling
efficiencies
  for various algorithms (MPI+thread: 1 MPI process and 24 threads per node)}}
  \label{fig_weakScaling}
\end{figure}

\major{Next, we evaluate,
the ability of our framework to process datasets of increasing sizes. For this, we}
\majorRevision{conduct a weak scaling analysis, where the
workload
increases
proportionally to 
the number of
available cores, starting at 24 cores (1 node).
The datasets have been resampled to $512^3$ on one node. For \emph{ScalarFieldCriticalPoints}, \emph{DiscreteGradient},
and \emph{ScalarFieldSmoother}, the input size is
increased
by doubling
the number of samples,
one dimension
at a time.
For \emph{IntegralLines}, the workload is increased by doubling the number of seeds at each
iteration.}
\majorRevision{Then,
we define the \emph{weak scaling efficiency}
for $p$ cores 
as $\frac{t_1}{t_p} 
\times 100$, with $t_1$ and $t_p$ being
the execution times on $1$ and $p$ 
nodes.
Appendix \discuss{C}
shows the same results as presented in Fig. \ref{fig_weakScaling}, but in terms 
of execution time instead of
 efficiency.}

\majorRevision{
As shown in Fig. \ref{fig_weakScaling}, 
for the \emph{ScalarFieldCriticalPoints} and the \emph{DiscreteGradient}, the efficiency
remains quite high as the amount of work and the number of cores double:
this is close to the ideal performance. Therefore, the conclusions are the same as for the strong scaling study:
the performance is very good on all data sets, slightly less for the \emph{DiscreteGradient} than the \emph{ScalarFieldCriticalPoints}.
For the \emph{ScalarFieldSmoother}, the weak scaling shows
that after the first drop of performance from one
to two processes, due to synchronizations and communications that do
not occur on one node, the computation actually scales really well, with
a nearly constant efficiency on more than one node.}

\majorRevision{For the \emph{IntegralLines}, the datasets \emph{Backpack}, \emph{Elevation}
and \emph{Isabel} show
degraded
performance similarly to the strong scaling. However, the results for the \emph{Wavelet} and \emph{Random}
stay much closer to the ideal than for the strong scaling study. This can be explained by two factors.
First, unlike the case of the strong scaling study, the number of seeds per node in the weak study is constant and does not decrease.
\discuss{Hence,} the workload imbalance has a 
smaller
impact and does not
deteriorate the performance as much. Second, it is likely that the workload for the strong scaling study
becomes too small as the number of cores increases. This makes the relative cost of communications
and synchronizations very \discuss{important}. 
}

\major{Overall, this weak scaling
analysis shows that, for \emph{ScalarFieldCriticalPoints} and \emph{DiscreteGradient}, the weak scaling is close to ideal (i.e. a problem of growing size can be processed in constant time when increasing accordingly the number of cores). 
For \emph{ScalarFieldSmoother}, after a first degradation due to inter-process synchronization and communication, the efficiency is nearly constant. 
Finally, weak scaling performances are degraded overall for \emph{IntegralLines}, at the exception of well balanced datasets that show much better performance than in the strong scaling study.}

\subsection{Integrated pipeline performance}
\label{sec_pipeline}

We now present experimental results for the integrated pipeline
(\autoref{sec_integratedPipeline}), which exemplifies a real-life use case
combining all of the port examples described in
\autoref{sec_exampleImplementations}, \major{on datasets which were too large
(8.5 and 120 billion vertices, \autoref{sec_integratedPipeline}) to be handled by TTK
prior to this work.}

\eve{The results \julien{for} the integrated pipeline are twofold: an output image (\autoref{fig_atExample} and \autoref{fig_dnsExample}) and the time profiling of the pipeline (\autoref{fig_profile})\julien{.}
The image is produced using offscreen rendering with OSMesa on \julien{our} supercomputer.
Profiling is done using both Paraview's timer
(\julien{average, minimum and maximum computation times across processes,}
for an overall
algorithm, preconditioning included) and the TTK timer defined in \autoref{sec_lowLevelInterface} 
(for a
\julien{fine-grain}
account
of the execution time within an algorithm and its preconditioning).}

\subsubsection{\eve{The Adenine Thymine complex (AT) dataset}}

\begin{figure*}
  \centering
  \includegraphics[width=\linewidth]{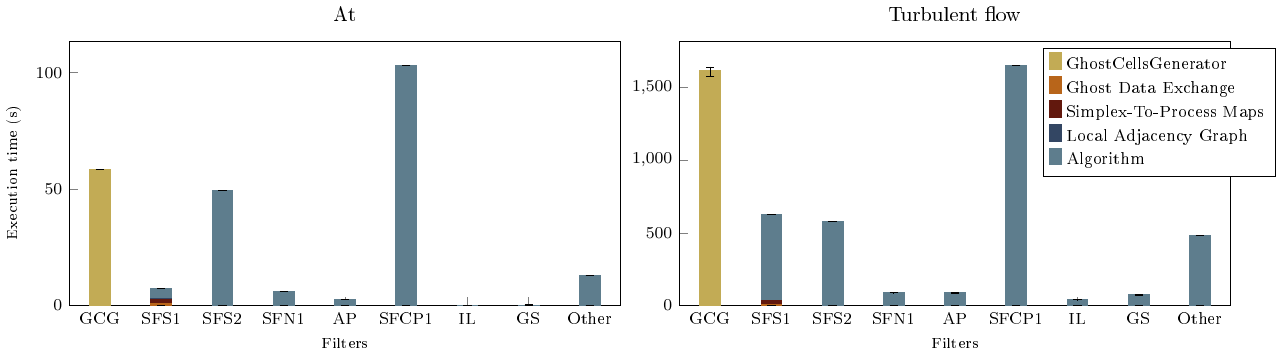}
  \caption{\eve{\julien{Time profiling}
  \julien{for}
  the integrated pipeline for the AT dataset
  resampled to roughly $8.5$ billion vertices
  (left) and the
  \emph{Turbulent Channel Flow}
  dataset (right) of 120 billion vertices. The execution
  was conducted using 64 nodes of 24 cores each 
  \majorRevision{($1536$}
  cores in total) on MeSU-beta.
  Each bar corresponds to the execution time of \julien{one} algorithm. SFS1 is computed for 1 iteration for the AT
  dataset and 10 iterations for the turbulent flow dataset \julien{(which} is more irregular).
  The \emph{Other} step consists in steps that are not part of an algorithm, such as loading the
  TTK plugin in Paraview, Paraview overhead and I/O operations. Only algorithms that take up a significant 
  amount of time are shown in the profiling \julien{(see} \autoref{table_integratedPipeline} for a description of the abbreviations\julien{)}. 
  \julien{In both cases, the MPI preconditioning
  computed by
  our framework
  (\emph{Local Adjacency Graph, \majorRevision{Simplex-To-Process Maps,}
Ghost Data Exchange})
\major{is}
negligible
\major{within}
the overall
pipeline execution time \eve{(at most $1.2 \%$)}.}}}
  \label{fig_profile}
\end{figure*}

\julien{For the experiments of Figs.~\ref{fig_atExample} and \ref{fig_profile} (left), the}
\eve{selected resampling dimensions \julien{for the input regular grid} are \majorRevision{$2048^3$}, a choice explained in \autoref{sec_integratedPipeline}.}
\eve{The overall computation takes $241.2$ seconds.
\majorRevision{Preconditioning}
is triggered once, before executing the first TTK algorithm.
The longest preconditioning step
is Paraview's ghost cells generation (\julien{$24.2 \%$ of the total pipeline time}), a step commonly used in a distributed-memory setting, regardless of TTK. The preconditioning specific to TTK's use of MPI \julien{(i.e. \emph{Local Adjacency Graph}, \emph{\majorRevision{Simplex-To-Process Maps}}, \emph{Ghost Data Exchange})} is
\julien{significantly}
faster and takes only $1.2\%$ of the overall \julien{pipeline computation} time, \julien{which can be considered as negligible
next
to the
rest
of the pipeline}.}
\eve{TTK computations (preconditioning included) make up $70.1\%$
 of the total \julien{pipeline} computation\julien{, which can be considered as a satisfactory efficiency.}}

\subsubsection{The Turbulent Channel Flow dataset}
\label{sec_dnsExample}

 \begin{figure*}
    \centering
    \includegraphics[width=\linewidth]{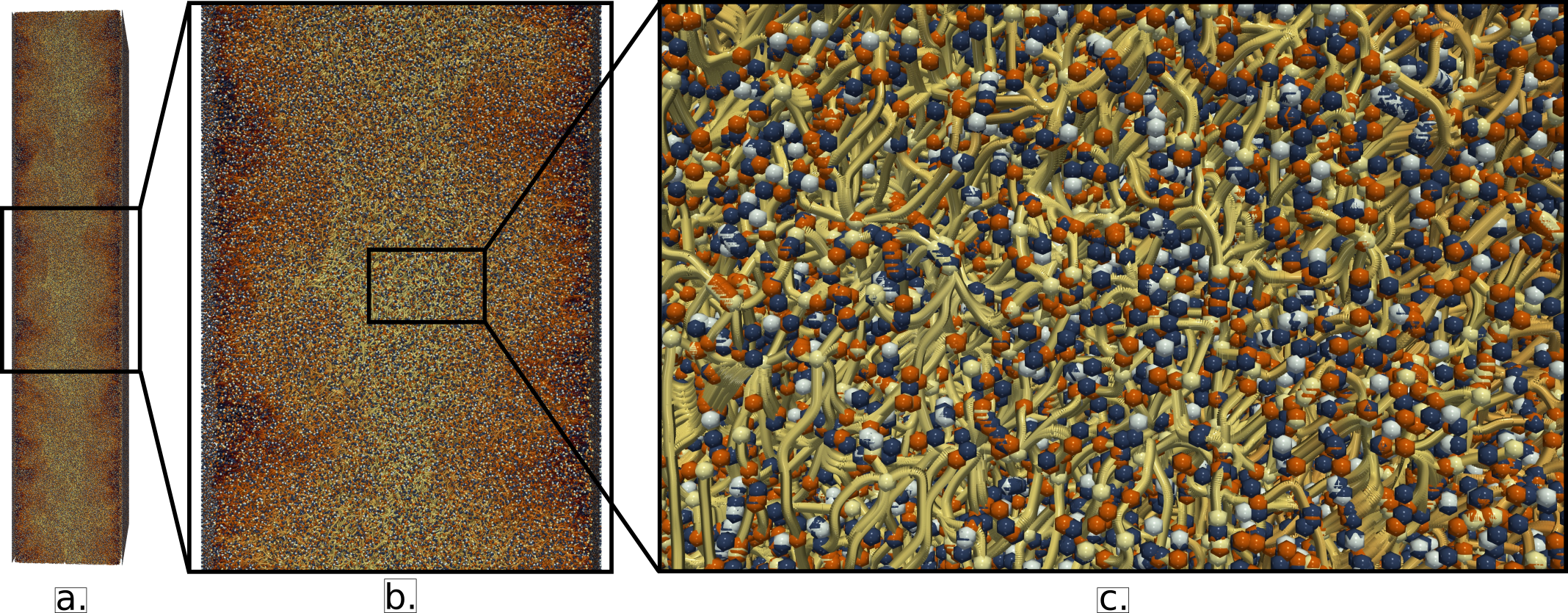}
    \caption{\eve{Output of the integrated pipeline on the
    \julien{\emph{Turbulent Channel Flow}}
    dataset (120 billion vertices), a three-dimensional
    regular grid with two scalar fields, the pressure of the fluid and its \julien{gradient} magnitude. The pipeline was executed up to the
    Geometry Smoother algorithm. The spheres \majorRevision{correspond} to the
    \julien{pressure}
    critical points and the tubes
    are the integral lines starting \majorRevision{at} saddle points.
    Figure (a) shows all of the produced geometry, while (b) and (c) show parts of the output zoomed
    in. These images were produced on a quarter of the total
    dataset due to rendering
    \julien{related}
    issues (\julien{see} \autoref{sec_dnsExample}),
    while \autoref{fig_profile} was produced on the full dataset.}}
    \label{fig_dnsExample}
  \end{figure*}

\eve{The computation shown in \autoref{fig_profile} \julien{(right)} was performed on the
\julien{complete}
dataset \julien{(120 billion vertices, single-precision, \autoref{sec_integratedPipeline}).}
The overall computation takes 
\majorRevision{$5257.5$}
seconds.
}

\eve{The execution time of this pipeline includes the algorithms listed in \autoref{table_integratedPipeline}. \julien{Note that the rendering time is not included in the time profiling reported in \autoref{fig_profile} (for both datasets).}
\julien{For the turbulent flow dataset, explicit}
 glyphs were
\julien{used for}
 the rendering \julien{of the critical points (spheres) and integral lines (cylinders),}
 as the screen-space glyph rendering features of ParaView did not produce satisfactory results in a distributed setting.
 \julien{However, the} generation of glyph geometry required a lot of memory, therefore the rendering in \autoref{fig_dnsExample} was performed
 on only a quarter of the dataset. The pipeline profiled in \autoref{fig_profile}, however, was indeed executed on the whole dataset.}

\eve{\julien{Similarly to the AT dataset, the}
longest preconditioning step
is Paraview's ghost cells generation \julien{($30.7 \%$ of the total pipeline time)}.
\julien{Again, TTK's specific MPI-preconditioning is marginal and takes up only $0.7 \%$ of the overall pipeline computation time.}
Computations of TTK algorithms (preconditioning included) make up for $59.2\%$ of the total execution time.
\eve{When compared to the AT dataset, the execution time of SFCP1 is multiplied by a factor of roughly 15, 
which is comparable to the increase in data size between datasets, indicating good scalability.}}

\julien{Overall, this experiment shows that, thanks to our MPI-based framework, TTK can now
run advanced analysis pipelines on
massive datasets (up to $120$ billion vertices on our supercomputer),
\major{which were too large to be handled by TTK prior to this work.
  We showed that this could be achieved
\discuss{in an acceptable amount of time},} 
while requiring a TTK-MPI specific preconditioning of negligible computation time overhead ($0.7 \%$ of
the
total
computation).}

\subsection{Limitations}
\julien{\autoref{sec_distributedTriangulation} presented our strategy to provide consistent global simplex identifiers, irrespective of the number of processes. This guarantees a per-bit compatibility of the input data representation with the sequential mode of TTK, and consequently a per-bit compatibility of the pipeline outputs.
However, the usage of threads can challenge the determinism of certain algorithms, given the non-deterministic nature of the thread scheduler. Then, an 
additional
effort may need to be made by the developers 
to address this non-determinism within their implementation
of a topological algorithm
(to ensure per-bit compatibility).}
\eve{In our experiments, we opted not to enforce determinism \julien{for integral lines,} 
given
the lack of control
over the thread scheduler.}
%
%


\eve{\julien{A significant difficulty occurring when processing massive datasets with ParaView is the substantial memory footprint induced by ParaView's interactive pipeline management.
Data}
flows through the pipeline, being transformed at each step by algorithms.
Rather than modifying data in-place, algorithms generate copies before implementing changes.
This methodology offers several advantages, such as preventing redundant computation of inputs
when multiple branches share the same input, resulting in better efficiency, especially when
\julien{adjusting interactively the algorithm parameters.}
However, this \julien{copy-before-computation} approach leads to a rapid increase in memory usage
during computations\julien{, which can become problematic in practice for pipelines counting a large number of algorithms.}}

\majorRevision{Finally, several specialized domain representations which are popular in scientific computing
-- such as grids with periodic conditions along a restricted set of dimensions or adaptive mesh refinement (AMR)  -- are not natively supported by TTK and these currently need to
be explicitly triangulated in a pre-process. In the future, we will investigate extensions of our distributed triangulation to support these representations natively, without pre-process.}

%% file: conclusion.tex
\section{Conclusion and Roadmap}
\label{sec_conclusion}

\eve{In this paper,
\julien{we presented a software framework for the support of topological analysis pipelines in a distributed-memory model. Specifically, we instantiated our framework with the MPI model, within the Topology ToolKit (TTK).}
An extension of TTK's efficient triangulation data structure
to a distributed-memory context was presented, as well as a software infrastructure supporting
advanced and distributed topological pipelines. A taxonomy of algorithms supported by TTK
was provided, depending on \julien{their} communication requirements. The ports of several algorithms were described,
with detailed performance analyses, following a MPI+thread strategy. We also provided a
real-life use case consisting of an advanced pipeline of multiple algorithms, run on a
dataset of 120 billion vertices on a compute cluster with 64 nodes
\majorRevision{($1536$}
cores), showing
that the cost of TTK\julien{'s MPI} preconditioning is marginal
\julien{next}
to the execution time of the pipeline.
TTK is now able to compute
complex pipelines involving several algorithms on datasets too \julien{large} to be processed on a commodity computer.
\julien{Our framework is available}
in TTK 1.2.0, enabling others to reproduce our results
or extend TTK's distributed capabilities.}
\majorRevision{Also, as TTK is now officially integrated in ParaView,
this distributed version of TTK will be available to a wide audience in the next
release of ParaView.}

\eve{The next step consists in adding distributed-memory support to all of TTK's
topological algorithms. The challenge here depends on the algorithm class
(see \autoref{sec_taxonomy}). The port of NC and DIC algorithms
(such as ContinuousScatterPlot, ManifoldCheck, DistanceField, JacobiSet or FiberSurface)
is relatively straightforward. For DIC algorithms, the initial step entails identifying the data to be exchanged, the processes
involved in the exchange, and the appropriate timing for these communications. For NC algorithms,
no exchange between processes take place.
Then, the implementation can be done in TTK, using TTK's MPI-API as well as low-level MPI directives
(for specific communications).
This could for example
be done during a hackathon. For DDC algorithms (such as Discrete Morse Sandwich,
Topological Simplification, Contour tree \major{or Rips complex computations}),
the port may be much more complicated.
For each of these DDC algorithms, their \pierre{distributed-memory parallelization} may be a substantial
research problem, on which we will focus in future work.}

%% file: appendixBody.tex
\appendices

\input{ttkBackground.tex}
\input{timeMeasurement.tex}
\input{performanceResultsAppendix.tex}

%% file: ttkBackground.tex
\section{\majorRevision{Background on The Topology ToolKit (TTK)}}

\major{This section provides some background regarding  the software library
\emph{the Topology ToolKit} (TTK), and it details its
pre-existing support (i.e. prior to this work) for triangulation traversal and
parallelization.}

\subsection{\major{Scope and interfaces}}
\major{TTK is an open-source software library for topological data analysis and
visualization, written in C++. While TTK can be used directly via
its raw, low-level C++ interface, TTK also provides an
interface of higher-level, for the \emph{Visualization ToolKit} (VTK
\cite{Kitware:2003}, another open-source C++ library, dedicated to data
visualization and analysis, but with a broader scope than TTK). In particular,
as described in its companion paper \cite{ttk17}, each TTK algorithm is
\emph{wrapped} into a VTK \emph{filter} (i.e. an elementary data processing
unit
in the VTK terminology).}
\major{Specifically, each topological algorithm implemented in TTK inherits
from the generic class
named} \verb|ttkAlgorithm|\major{, itself inheriting from the generic VTK data
processing class named} \verb|vtkAlgorithm|\major{.
Then, when reaching a TTK algorithm within a distributed pipeline,
ParaView will call the
function} \verb|ProcessRequest| \major{(from the}
\verb|vtkAlgorithm| \major{interface, see Fig. 6).
The re-implementation of this function in the} \verb|ttkAlgorithm| \major{class
will trigger all the necessary preconditioning
before calling the actual topological algorithm (see Sec. 6
for examples), implemented in the generic function}
\verb|RequestData| \major{(from the} \verb|vtkAlgorithm|
\major{interface).}

\major{Thanks to this \emph{wrapping},
a developer can use TTK features with
the same syntax as VTK features. TTK also provides a plugin for the open-source
application \emph{ParaView} \cite{paraviewBook}, which is a de-facto standard
for the visualization and analysis of large-scale data. Then, ParaView users
can interactively call TTK filters via its graphical user interface. Finally,
TTK also provides two Python interfaces (a low-level one, matching its VTK
interface, and a high-level one, matching its ParaView interface).}

\subsection{\major{Pre-existing triangulation}}
\major{Internally, each topological algorithm implemented in TTK is exploiting
TTK's
generic data-structure
for the efficient traversal of simplicial complexes (Sec. 4.1), a central
aspect in most
topological algorithms.
All
traversal queries (e.g. getting the $i^{th}$ $d''$-dimensional co-face of a
given $d'$-simplex $\simplex$) are addressed by the data structure in constant
time, which is of paramount importance to guarantee the runtime performance of
the calling topological algorithms. This is supported by the data structure via
a preconditioning mechanism. Specifically, in a pre-processing phase, each
calling topological algorithm needs to explicitly declare the list of  the
types of traversal queries it is going to use during its main routine. This
declaration will trigger a preconditioning of the triangulation, which will
pre-compute and cache all the specified queries, whose results will
 later be addressed in constant time at query time.
This design philosophy is particularly relevant in the context of analysis
pipelines, where multiple algorithms are typically combined together. There,
the preconditioning phase only pre-computes the information once (i.e. if it is
not
already available in cache). Thus, multiple algorithms can benefit from a common
preconditioning of the data structure. Moreover,
another benefit of
this
strategy is that it
adapts the memory footprint of the data structure, based on the types of
traversals required by the calling algorithm.}

\major{In the specific case of regular grids, adjacency
relations
can be easily inferred, given the regular pattern of the grid sampling
(considering the Freudenthal
triangulation \cite{freudenthal42, kuhn60} of the grid). Then,
TTK's triangulation
supports an \emph{implicit}
mode for regular grids: for such inputs, the preconditioning does not
store any information and the results of all the queries are
\majorRevision{computed on-the-fly} at
runtime \cite{ttk17}.
An extension to periodic grids (i.e. with periodic boundary
conditions\majorRevision{, for all dimensions})
is also implemented.
The switch from one implementation to the other (explicit
mode for meshes or implicit mode for
grids) is automatically handled by TTK and developers of topological algorithms
only need to produce one implementation, interacting with TTK's
generic triangulation data structure.
}

\subsection{\major{Pre-existing parallel algorithms (shared-memory)}}

\major{TTK implements a substantial collection of topological algorithms for
scalar
data, bivariate data, ensemble data or even point cloud data. For more
details, we refer the reader to an overview paper \cite{ttk19} as well as to
\emph{TTK's
Online Example Database} \cite{ttkExamples} (a database of real-life
data analysis use cases, implementing advanced topological analysis pipelines,
combining multiple algorithms).
}

\major{Prior to this work, only shared-memory parallelism was implemented in
TTK, \discuss{using multiple threads} with OpenMP \cite{openmp51}. Then parallel computations could only be
carried out on a single computer.
Specifically, some of the topological objects
introduced
in the manuscript
(the critical points and the discrete
gradient) could be computed in parallel (the integral line extraction
implementation however was sequential). TTK provides shared-memory parallel
computations for various objects, including the following, non-exhaustive list:
continuous scatterplots \cite{BachthalerW08a},
data or geometry smoothing,
dimensionality reduction \cite{topoMap},
fiber surfaces  \cite{KlacanskyTCG17},
Jacobi sets \cite{jacobiSets},
mandatory critical points \cite{mandatory},
marching tetrahedra,
merge and contour trees \cite{gueunet_tpds19, lukasczyk_vis23},
merge tree distances and encoding \cite{pont_vis21, wetzels_vis23, pont_tvcg22,
pont_tvcg24},
Morse-Smale complexes \cite{ttk17},
path compression \cite{robin23},
persistence diagrams \cite{guillou_tech22},
persistence diagram encoding \cite{sisouk_tvcg24},
Reeb graphs \cite{gueunet_egpgv19},
Reeb spaces \cite{tierny_vis16},
Rips complexes,
scalar field normalizer,
topological compression \cite{soler_pv18},
topological simplification \cite{tierny_vis12, Lukasczyk_vis20}.
}

\subsection{\major{Contributions}}

\major{In this work, we document the infrastructure evolution that
is required for TTK to support distributed parallelism, via MPI, as documented
in Secs.
3
(distributed data model),
4
(distributed mesh data-structure) and
5
(distributed pipeline management).
We also provide examples of topological algorithms
(Sec. 6)
which we extended to support distributed computations with MPI, on top of their
pre-existing shared-memory parallelization with OpenMP. In particular, note
that
Sec. 6.3.5
describes a shared-memory parallelization with
OpenMP of integral line computation which is novel in this work (this
computation was sequential in the pre-existing version of TTK). Finally, we
provide in
Sec. 8
a roadmap for the extension to the
distributed computation of the remaining algorithms of TTK.
}

%% file: timeMeasurement.tex
\noindent
\section{\majorRevision{Fine-scale time performance measurements}}
\label{appendix_timeMeasurement}
\majorRevision{
 When timing the execution of a specific distributed algorithm,
simply measuring the execution time on one process may
not 
represent the execution time of the whole algorithm, as
the local execution time may greatly vary from one process to
  the next.
An established way to measure time in a distributed-memory environment consists
in adding
a MPI barrier 
before starting and stopping the timer. The call before
starting the timer forces all processes to start simultaneously and the call
before stopping the timer ensures that the time measurement includes the slowest
process. Doing so, the execution time from (e.g.) process 0
 then corresponds to the overall MPI execution time. In TTK, this can be done using the two functions} \verb|startMPITimer|
\majorRevision{and} \verb|stopMPITimer|.

\majorRevision{
 However, the two MPI barriers add synchronization points, slowing down
 the execution. Hence, the execution time is not measured
by default but only when the compilation variable}
 \verb|TTK_ENABLE_MPI_TIME| \majorRevision{is set to} \verb|ON| \majorRevision{for TTK.}


%% file: performanceResultsAppendix.tex
\label{appendix_strongScaling}
\begin{figure}
  \centering
  \includegraphics[width=\linewidth]{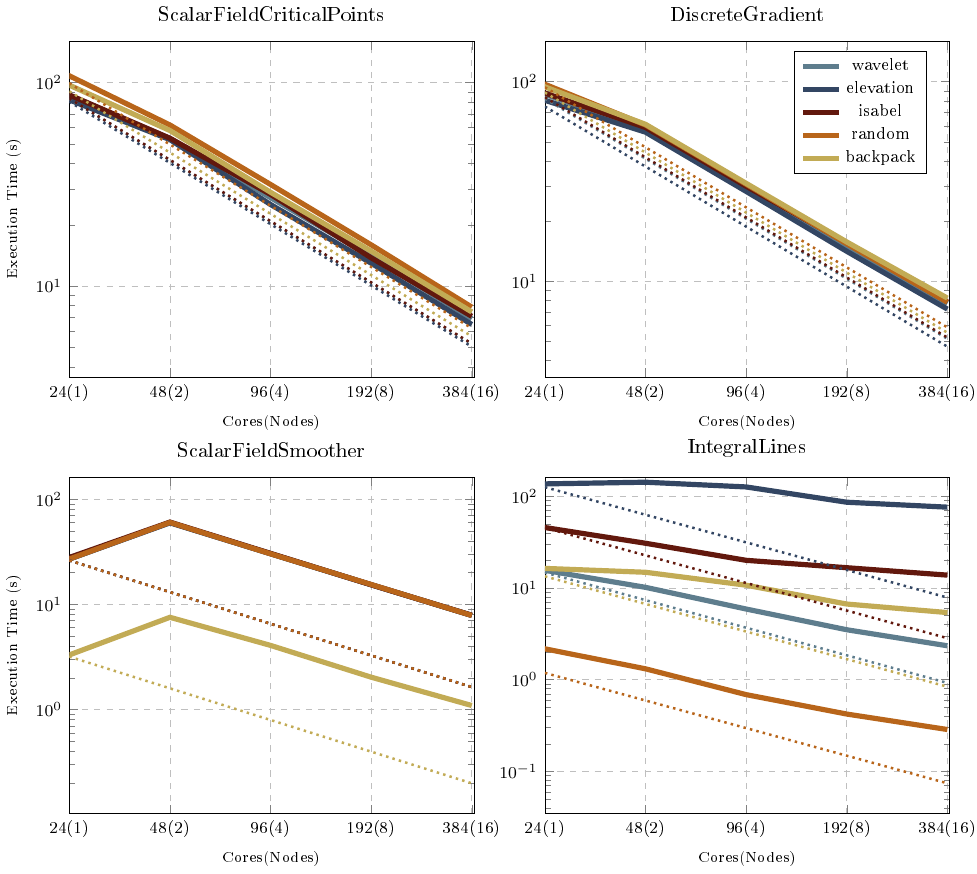}

  \caption{\majorRevision{
  Strong scaling (execution times)
    for various
algorithms
        \major{(MPI+thread: 1 MPI process and 24 threads per node). The
dotted lines indicate ideal performances.}}}
  \label{fig_strongScaling}
\end{figure}

\begin{figure}
  \centering
  \includegraphics[width=\linewidth]{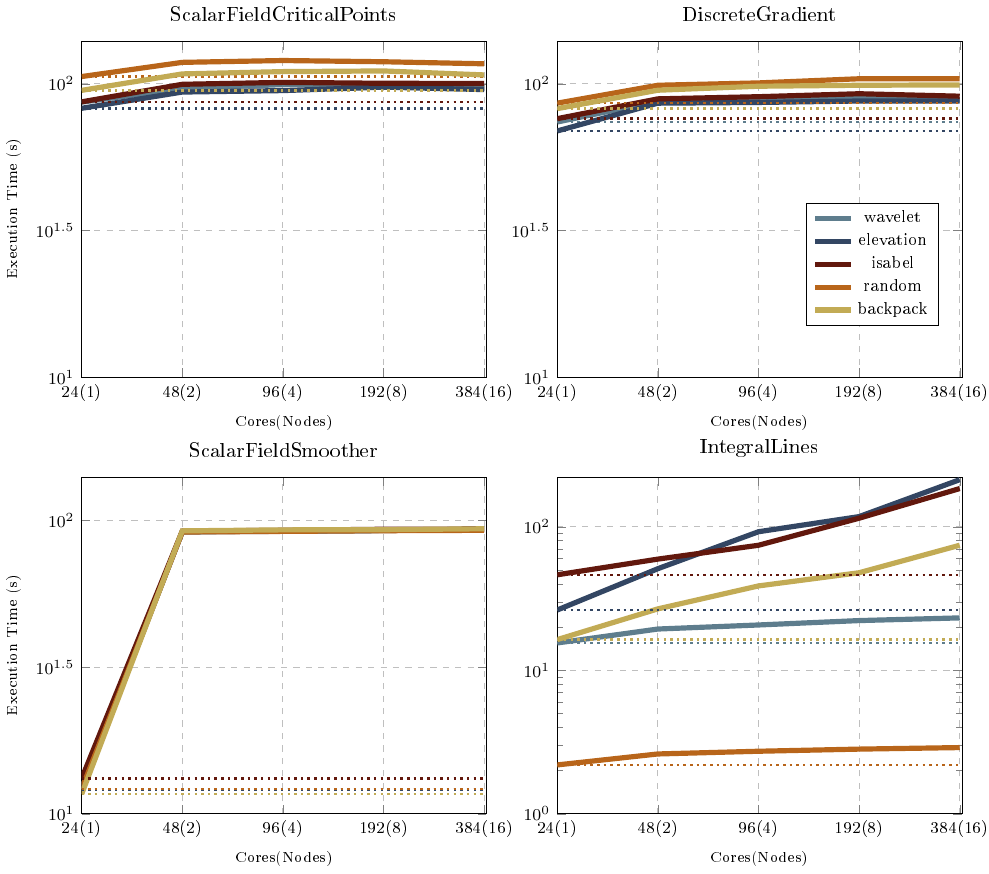}
  \caption{\majorRevision{Weak scaling
(execution times)
  for various algorithms (MPI+thread: 1 MPI process and 24 threads per node). The
  dotted lines indicate ideal performances.}}
  \label{fig_weakScalingTime}
\end{figure}

\section{\majorRevision{Additional Performance Results}}

\majorRevision{This appendix provides further details regarding the performance of the
distributed algorithms presented in this paper.}

\major{Specifically, \autoref{fig_strongScaling} and \autoref{fig_weakScalingTime} show the same performance
results as in the section 7.1.1 and section 7.1.2 of the main manuscript (\emph{``Strong
scaling''}) and (\emph{``Weak scaling''}), but expressed in terms of running time instead of efficiency.}
